\documentclass[11pt,a4paper]{article}

\usepackage{jcappub}
\usepackage[margin=0.95in]{geometry}

\usepackage[utf8]{inputenc}
\usepackage{graphicx}
\usepackage{subcaption}
\usepackage[export]{adjustbox}
\usepackage{amsmath}
\usepackage{hyperref}

\renewcommand\theequation{\arabic{section}.\arabic{equation}}

\newcommand{\nn}{\nonumber}

\newcommand{\llp}{\left [}
\newcommand{\rrp}{\right ]}

\newcommand{\lp}{\left (}
\newcommand{\rp}{\right )}

\def\PBH{\text{\tiny  PBH}}

\newcommand{\be}{\begin{equation}\begin{aligned}}
\newcommand{\ee}{\end{aligned}\end{equation}}
\newcommand{\bbe}{\begin{align}}
\newcommand{\eee}{\end{align}}
\newcommand{\bea}{\begin{eqnarray}}
\newcommand{\eea}{\end{eqnarray}}
\def\beq{\begin{equation}}
\def\eeq{\end{equation}}

\def\lsim{\mathrel{\rlap{\lower4pt\hbox{\hskip0.5pt$\sim$}}
		\raise1pt\hbox{$<$}}}     
\def\gsim{\mathrel{\rlap{\lower4pt\hbox{\hskip0.5pt$\sim$}}
		\raise1pt\hbox{$>$}}}     
	
	\def\d{{\rm d}}	
\newcommand{\arXiv}[2]{\href{http://arxiv.org/pdf/#1}{{\tt [#2/#1]}}}
\newcommand{\arXivold}[1]{\href{http://arxiv.org/pdf/#1}{{\tt [#1]}}}

\title{The  Astro-Primordial Black Hole Merger Rates: a Reappraisal}
\author[a]{K. Kritos,}
\author[b]{V. De Luca,}
\author[b]{G. Franciolini,}
\author[a]{A. Kehagias,}
\author[b,c]{A. Riotto}

\affiliation[a]{Physics Division, National Technical University of Athens, Zografou, Athens, 15780, Greece} 
\affiliation[b]{Department of Theoretical Physics and Center for Astroparticle Physics (CAP) \\
			24 quai E. Ansermet, CH-1211 Geneva 4, Switzerland}
\affiliation[c]{INFN, Sezione di Roma, Piazzale Aldo Moro 2, 00185, Roma, Italy}

\abstract{
	Mainly motivated by the   recent GW190521 mass gap event which we take as a benchmark point, we  critically assess if    binaries made of a primordial black hole and a black hole of astrophysical origin may  form, merge in stellar clusters and reproduce the LIGO/Virgo detection rate.
While two previously studied mechanisms 
-- the direct capture and the three body induced -- seem to be inefficient, we propose a new ``catalysis'' channel based on the idea that a subsequent chain of  single-binary and binary-binary exchanges may lead   to the formation of a high mass binary  pairs and  show that it  may explain  the recent GW190521 event if the local overdensity of primordial black holes  in the globular cluster is larger than a few.
}
\emailAdd{ge16004@central.ntua.gr}
\emailAdd{valerio.deluca@unige.ch}
\emailAdd{gabriele.franciolini@unige.ch}
\emailAdd{kehagias@central.ntua.gr}
\emailAdd{antonio.riotto@unige.ch}
\begin{document}

\maketitle
\flushbottom

\section{Introduction}
\noindent
Since the very first detection of a compact object merger by the Laser Interferometer Gravitational Wave Observatory (LIGO) \cite{TheLIGOScientific:2014jea}, it was realized that the Primordial Black Hole (PBH) scenario is an intriguing possibility that could potentially explain the origin of the events \cite{Bird:2016dcv,Sasaki:2016jop, Blinnikov:2016bxu, Clesse:2016vqa}. PBH-PBH mergers could form in the early or in the late universe \cite{Nakamura:1997sm,raid,Raidal:2017mfl,Ali-Haimoud:2017rtz, Inman:2019wvr} and their merger rates and properties have been studied in detail \cite{Vaskonen:2019jpv,Gow:2019pok,Liu:2019rnx,DeLuca:2020qqa,Jedamzik:2020ypm,Jedamzik:2020omx, DeLuca:2020jug,Tkachev:2020uin, Hall:2020daa, Wong:2020yig, Dolgov:2020xzo,Clesse:2020ghq, Hutsi:2020sol}.  

During the third observing run the LIGO, Virgo and KAGRA collaboration 
has released several detections of binary black hole (BH) mergers \cite{Abbott:2020niy,Abbott:2020gyp}. Within this catalog, probably the most noteworthy event is the one with at least one object in the so called BH mass gap, i.e. GW190521 \cite{Abbott:2020mjq}. In particular, the GW190521 primary mass confidently lies in the upper mass gap ranging from $\simeq65 M_\odot$ to $\simeq120 M_\odot$, which is associated to the pair-instability supernovae \cite{Abbott:2020tfl,Abbott:2020mjq}.
The formation of a GW190521-like merger may be naturally explained within the PBH scenario, see for example Ref.~\cite{DeLuca:2020sae}. Other possibilities rely on astrophysical formation scenarios. For example, the members of such high mass mergers could be the result of a runaway process of merging of lower mass BHs inside dense stellar clusters \cite{Liu:2020gif,Anagnostou:2020umw,Kimball:2020qyd,Fragione:2020han,Gerosa:2019zmo,Mapelli:2020xeq, Baibhav:2020xdf,Antonini:2018auk, Rodriguez:2019huv} or they could arise via multiple stellar coalescences \cite{Kremer:2020wtp}, or from population III  remnants \cite{Liu:2020lmi,Kinugawa:2020xws,Safarzadeh:2020vbv, Belczynski:2020bca}. Regarding the assembly of such an event, it might be a highly eccentric merger \cite{Gayathri:2020coq,Romero-Shaw:2020thy,Gondan:2020svr}, an asymmetric coalescence \cite{Fishbach:2020qag, Nitz:2020mga}, or it could have been formed in the environment of an active galactic nuclei \cite{Samsing:2020tda,Tagawa:2020jnc}.

If PBHs constitute a sizeable fraction of the dark matter (DM)  (for a recent comprehensive review on current constraints see \cite{Sasaki:2018dmp, Carr:2020gox, green}), then they will most likely populate astrophysical systems, like stellar clusters. In these environments a population of BHs that originate from stellar evolution, henceforth Astrophysical BHs (ABHs) is hosted as well. The rate of ABH-ABH mergers from such environments has also been surveyed in the past \cite{Rodriguez:2015oxa,Rodriguez:2016kxx,OLeary:2005vqo,Miller:2008yw,Antonini:2016gqe}. Given this likely coexistence of PBHs and ABHs in the same system, we expect that PBHs will be interacting with ABHs and forming binaries with them that might subsequently coalesce. Therefore, PBH-ABH mergers should be inevitable. Because PBHs were formed in the early universe and ABHs arose from stellar evolution in the late universe, PBH-ABH mergers can only be assembled via some dynamical channel, for example through a capture process \cite{Nitz:2020bdb,Tsai:2020hpi,Vattis:2020iuz}.

In this work, motivated by the recent GW190521 event which we take as a benchmark point, we  wish to critically assess if   PBH-ABH binaries can form, merge in stellar clusters and reproduce the LIGO/Virgo detection rate.
Specifically, we evaluate the contribution from the Milky Way type globular clusters in calculating the merger rate. This has also been done in the past, for instance in \cite{Kovetz:2018vly,Kritos:2020fjw} where it is estimated that there is a non-negligible contribution from globular clusters.
For simplicity we take a monochromatic mass spectrum for the PBHs peaked at the value $80M_\odot$. This is motivated by the necessity to  have PBHs well inside the upper mass gap where current constraints allow for  a total PBH abundance sufficient to obtain sizeable merger rates. Notice also that the choice of a broader PBH mass spectrum, e.g.  a log-normal shape typically predicted by PBH formation scenarios, is not expected to modify our results if narrow enough.
Furthermore, we take an illustrative doubly peaked spectrum for the ABHs with a low mass and a high mass component at $10M_\odot$ and $60M_\odot$, respectively. This choice  regarding the astrophysical mass spectrum is admittedly oversimplifying, but it allows to deal with an otherwise complex dynamical system. We consider this as the first step towards a more thorough analysis.

We look at three  possible formation mechanisms: direct capture, three body induced processes and a runaway process of binary-single and binary-binary exchanges leading to the formation of a high mass ABH-PBH pair. To investigate this third case which we refer to as the ``catalysis'' channel,  we 
resort to meaningful analytical estimates, comparing the rates and timescales of various processes occurring inside the clusters, in order to critically test whether ABH-PBH binaries can efficiently form.

The paper is organized as follows.  In Sec.~\ref{prelims_section} we make necessary definitions and present a few preliminary concepts. Then, in Sec.~\ref{DC_section} we focus on the direct capture and in Sec.~\ref{3ind_section} on three body induced channels. In Sec.~\ref{catalysis_section} we develop the ``catalysis'' model and discuss our findings. Finally, in Sec.~\ref{conclusions_section} we provide some concluding remarks.

As the paper will contain several acronyms, in the following Table and for the reader's sake we summarise a list of abbreviations used throughout this work.

\vskip 0.5cm
\begin{table}[h]
	\centering
	\begin{tabular}{c|c}
		\hline
		\hline
		BH & Black Hole \\
		ABH & Astrophysical Black Hole \\
		PBH & Primordial Black Hole \\
		DM & Dark Matter \\
		GW & Gravitational Wave \\
		GC & Globular cluster \\
		MWGC & Milky Way Globular Cluster \\
		DC & Direct Capture \\
		SMA & Semi-Major Axis \\
		3-ind. & three body induced binary \\
		3rd & third body binary hardening \\
		\hline
		\hline
	\end{tabular}
	\caption{List of abbreviations used in this work.}
	\label{table_abbr}
\end{table}

\section{Preliminaries}
\label{prelims_section}
\noindent
In this work we focus on the Milky Way globular clusters (MWGCs) and 
adopt the MWGC parameters inferred using the data from the catalog provided in Ref.~\cite{harris2010new}\footnote{The catalog is available online at \href{https://www.physics.mcmaster.ca/~harris/mwgc.dat}{https://www.physics.mcmaster.ca/~harris/mwgc.dat}.}, as was adopted recently in \cite{Kritos:2020fjw}.
In particular, the catalog provides information on the masses, concentrations, central densities and size of the baryonic mass distribution of the clusters.
We assume a King profile \cite{King1962} for the mass distribution of stars in each globular cluster (GC). For the normalization of the mass density profile we take the value of the profile at $r=0$ to match the central density value as provided by the catalog in Ref.~\cite{harris2010new}. We denote by $M_\text{\tiny h}$ the half-mass contained within the half-mass radius $r_\text{\tiny h}$ of the cluster. We also write $M_\text{\tiny cl}=2M_\text{\tiny h}$ for the cluster's total mass. 
 For simplicity, we keep into account the evolution of clusters through time in an effective way by fixing the fraction of retained BHs in binaries in order to match the results of the numerical simulation performed in Ref.~\cite{Chatterjee:2016hxc}, see Sec.~\ref{catalysis_section} for details. Furthermore, we assume all GCs to have a lifetime of $T_\text{\tiny cl} \simeq10$ Gyr \cite{Kritos:2020fjw,Valcin:2020vav,Flitter:2020bky}, and neglect the spin of the BHs in the analysis. The number density of globular clusters (GCs) in the local universe is estimated using the conventional value of $n_\text{\tiny GC}=2.4^{+0.9}_{-0.9}\,\text{Mpc}^{-3}$ \cite{Rodriguez:2016kxx,PortegiesZwart:1999nm,Antonini:2020xnd,Kremer:2020wtp,Flitter:2020bky}. In any case, our final results for the merger rate density will scale proportional to this number density.

It turns out that the merger rate considering dynamical scenarios in stellar clusters depends strongly on the structural characteristics of these environments. Specifically, the density, velocity and total number of species (ABHs and PBHs) that are merging affect significantly our results for the merger rates. For instance,  low density environments or high velocity dispersions results in small capture rate between two BHs. For these reasons, in the rest of this section we will review all of the above in the context we are interested in.

\subsection{Velocity dispersion and segregation radii}

The mean rms velocity of stars inside a cluster in terms of the cluster's half-mass and half-mass radius is given by (see  Ref.~\cite{1971ApJ...164..399S})
\begin{equation}
\label{sstar}
\sigma_*=\sqrt{{4\over5}{GM_\text{\tiny h}\over r_\text{\tiny h}}}\simeq19\, \text{km/s}\lp{M_\text{\tiny h}\over10^5M_{\odot}}\rp^{1/2}\lp{r_\text{\tiny h}\over \text{pc}}\rp^{-1/2},
\end{equation}
which represents the three dimensional velocity dispersion of stars with a mean mass $\langle m\rangle \simeq M_\odot$\footnote{Taking the mean mass of the stars to be different from $1 M_\odot$ by a factor of order unity would not alter our results.}. The escape velocity would simply be given by
$V_\text{\tiny esc}=2\sqrt{3}\sigma_*$ \cite{Fragione:2020fnr}.
Heavier objects like BHs interact with the stars inside the cluster and their velocity is relaxed to a smaller value relative to $\sigma_*$ through dynamical friction \cite{1987gady.book.....B}. Defining $m$ as the mass of the object,  then
\begin{equation}
\sigma(m)=\sqrt{{\langle m\rangle\over m}K\left(m,\langle m\rangle\right)}\ \sigma_*,
\label{sigma_m}
\end{equation}
where $K(m,\langle m\rangle)$ is the mass dependent ratio of the kinetic energies whose value can be numerically determined.
The typical values of  $K(m,\langle m\rangle)$ lie within ${\cal O}(1 \div 10)$ depending on the masses under consideration \cite{Khalisi:2006ne} (larger values obtained for larger hierarchies between $m$ and $\langle m \rangle$).
Indeed, it was demonstrated in Ref.~\cite{Khalisi:2006ne} that energy equipartition is rarely achieved inside of clusters when $m>2\langle m\rangle$. In this work we use $K(m,\langle m\rangle)\simeq(m/2\langle m\rangle)^{0.68}$ which is based on the data in Fig.~15 of \cite{Khalisi:2006ne}, and the corresponding dependence of the velocity dispersion in terms of mass goes like $\sigma(m)\propto m^{-0.15}$, which is consistent with the scaling found in Ref.~\cite{Trenti:2013ety}  for GCs. Therefore, the heaviest objects in the cluster tend to sink towards the center and segregate  within a sphere of radius $R_\text{\tiny seg}(m)$ determined by Eq.~\eqref{sigma_m} which, when combined with the Virial theorem and Eq.~\eqref{sstar}, results in an algebraic equation 
\begin{equation}
{M(R_\text{\tiny seg}(m))\over M_\text{\tiny h}}={\langle m\rangle\over m}K(m,\langle m\rangle){R_\text{\tiny seg}(m)\over r_\text{\tiny h}}.
\label{find_Rm}
\end{equation}
Here, $M(r)$ is the total mass contained within a sphere of radius $r$ from the center of the cluster and is found by integrating the King profile within that sphere. We find by direct numerical solution of Eq.~\eqref{find_Rm} that the segregation radius for BH masses used in this work is smaller than the core radius, and therefore we take the BHs and binary BH populations to be uniformly distributed within their segregation volume.\footnote{We do not consider here the possibility that an intermediate mass BH is located at the very center of the GC, which might induce a cusp in the BH distributions \cite{Bahcall:1976aa}.}

Finally, we give the expression for the relative velocity between two objects $A$ and $B$ in the cluster with masses $m_A$ and $m_B$ respectively. Assuming that on average the objects approach at right angles, then we have
\begin{equation}
\sigma_{A,B}=\sqrt{\sigma(m_A)^2+\sigma(m_B)^2}=\left[\lp{K(m_A,\langle m\rangle)\over m_A}+{K(m_B,\langle m\rangle)\over m_B}\rp\langle m\rangle\right]^{1/2}\sigma_*.
\end{equation}

\subsection{Primordial and astrophysical black holes inside stellar clusters}

We assume that a population of PBHs is already hosted inside the  cluster. We parametrize the abundance of PBHs by two fractions. First, we account for the global fraction of DM in the form of PBHs, $f_\text{\tiny PBH}$. 
Motivated by the recent GW190521 event, we consider a monochromatic mass spectrum of PBHs peaked at the mass of $80M_{\odot}$. This choice is analogous to the one adopted in Ref.~\cite{DeLuca:2020sae} in their single population PBH scenario possibly giving rise to GW190521 as a PBH-PBH binary.  

Second, we parametrise the uncertainty on the local density of PBHs within the GC by  introducing the parameter  $f_\text{\tiny 80}$ which characterises the abundance of PBHs with respect to the baryonic matter estimated by assuming that the GC has a DM mass comparable to the baryonic mass $M_\text{\tiny cl}$.  Therefore, the number of PBHs hosted inside our clusters is parametrised as 

\begin{equation}
N_\text{\tiny 80}=f_\text{\tiny 80}\left({M_\text{\tiny cl}\over80M_{\odot}}\right).
\label{N_80}
\end{equation}
We are assuming PBHs as isolated objects and not bounded in PBH binaries. This is expected to be an accurate assumption as only a small portion of the PBH population participates in early universe binaries formed before matter-radiation equality.
One can estimate this quantity by computing the probability that two PBHs are located within a distance small enough to allow for a decoupling of the system from the Hubble flow~\cite{Nakamura:1997sm,raid,Ali-Haimoud:2017rtz}. By using the initial Poisson distribution for the spatial PBH correlation at high redhsifts \cite{cl1,cl2,cl3,cl4},  the fraction of PBHs in binaries is therefore 
\begin{equation}\label{fracPBHinbin}
f^\text{\tiny binaries}_{\text{\tiny PBH}}
\simeq f_\PBH^2/2.
\end{equation}
For the parameter space we consider in the following, we find this fraction to be negligible and the initial assumption of isolated PBHs in GCs is justified.

For ABHs we use the Kroupa initial mass function to estimate the fraction of the cluster's total mass that is in the form of ABH progenitors \cite{Kroupa:2002ky}. This fraction integrates to approximately $10\%$ for star progenitor masses in the range from $20M_\odot$ to $120M_\odot$. The ABH mass fraction should be reduced by a factor of order three which accounts for the fact that only a fraction of the progenitors mass corresponds to the BH remnant \cite{Fryer:1999ht}. We denote the fraction of the cluster's mass associated with ABHs by $f_\text{\tiny ABH}\simeq{1/3}\times10\%\simeq3\%$ (see also Eq.~(7) in \cite{Kritos:2020fjw}). For illustrative purposes, in this work we also simplify the ABH continuous mass spectrum and consider a simpler model with two monochromatic contributions corresponding to  $10M_{\odot}$ and $60M_{\odot}$,  respectively. The relative abundances $f_\text{\tiny 10}$ and $f_\text{\tiny 60}$ of these two populations in this two-mass model are estimated by constraining their sum to equal one and their ratio to follow the ratio of the corresponding ABH mass function, which is taken to be \cite{Kovetz:2016kpi}
\begin{equation}
	P(m_\text{\tiny ABH})\propto\Theta(m_\text{\tiny ABH}-M_\text{\tiny gap})m_\text{\tiny ABH}^{-2.35}\exp\left(-m_\text{\tiny ABH}/M_\text{\tiny cap}\right),
\label{P_BlackHoleMass}
\end{equation}
where  $M_\text{\tiny gap}=5M_{\odot}$, $M_\text{\tiny cap}=40M_{\odot}$ \cite{catalog} and  $\Theta$ denotes the Heaviside step function. Slightly different values for the upper exponential  cutoff mass scale $M_\text{\tiny cap}$ would not impact significantly the results presented in this paper.
Finally, one finds ${f_\text{\tiny 60}/ f_\text{\tiny 10}}\simeq 4.3\times10^{-3}	$.

We also need to account for the fact that ABHs receive a natal kick after birth and a fraction of them escapes the gravitational potential of the cluster, see Refs.~\cite{Chatterjee:2016hxc, Pavlik2018}. We assume the kick of the ABHs follows a Maxwellian distribution $P_\text{\tiny kick}(v_\text{\tiny kick})$ with one dimensional velocity dispersion of $w_\text{\tiny kick}$. The parameter $w_\text{\tiny kick}$ is treated as a free input in the computation as it is affected by large uncertainties related to the ABH formation process. The retention fraction $f_\text{\tiny ret}$ can be estimated by integrating the Maxwellian distribution up to the escape velocity $V_\text{\tiny esc}$, 
\begin{equation}
f_\text{\tiny ret}=\int_0^{V_\text{\tiny esc}}P_\text{\tiny kick}(v_\text{\tiny kick})dv_\text{\tiny kick}=\text{erf}\left({V_\text{\tiny esc}\over\sqrt{2}w_\text{\tiny kick}}\right)-\sqrt{2\over\pi}{V_\text{\tiny esc}\over w_\text{\tiny kick}}\exp\left[-\left({V_\text{\tiny esc}\over\sqrt{2}w_\text{\tiny kick}}\right)^2\right].
\end{equation}
In particular, for small enough values of the ratio $V_\text{\tiny esc}/w_\text{\tiny kick}$, the retention fraction scales like $f_\text{\tiny ret}\propto\lp V_\text{\tiny esc}/w_\text{\tiny kick}\rp^{2.98}$.
The corresponding maximum numbers of ABH populations inside each cluster are defined as 
\begin{align}
N_\text{\tiny 10}&=f_\text{\tiny 10}f_\text{\tiny ret}f_\text{\tiny ABH}\left({M_\text{\tiny cl}\over 10M_\odot}\right),\label{N_10}  \\
N_\text{\tiny 60}&=f_\text{\tiny 60}f_\text{\tiny ret}f_\text{\tiny ABH}\left({M_\text{\tiny cl}\over 60M_\odot}\right).
\label{N_60}
\end{align}
In Table~\ref{table_BHs} we estimate the integral number of BH populations in characteristic MWGCs and using the Eqs.~\eqref{N_80},~\eqref{N_10} and~\eqref{N_60}. We have considered two values for $w_\text{\tiny kick}= 50 \, {\rm km/s}$ and $w_\text{\tiny kick}= 100 \, {\rm km/s}$. As can be seen from Table~\ref{table_BHs}, the maximum number of ABHs drops quickly with increasing $w_\text{\tiny kick}$ indicating that GCs tend to retain a very small fraction of their ABHs if the natal kick is relatively high. As we will see in Sections~\ref{DC_section}, \ref{3ind_section} and \ref{catalysis_section}  this will significantly affect the ABH-PBH merger rate. Let us stress that this estimate does not take into account the subsequent evaporation of the globular clusters, which has a crucial impact on the retained number of BHs~\cite{Chatterjee:2016hxc}. As we will discuss  in Sec.~\ref{catalysis_section}, this effect has been captured in our computation by accounting for the proper number of retained BH binaries following the numerical simulation done in Ref.~\cite{Chatterjee:2016hxc}.
 For the last column, where we calculate the number of 80 $M_\odot$ PBHs, we have indicatively used $f_\text{\tiny 80}=10^{-4}$.  We motivate this choice by the following considerations. 
The overall PBH relative global abundance $f_\text{\tiny PBH}$ should be smaller than ${\cal O}(10^{-5})$ as for larger values the merger rate of PBH-PBH binaries would dominate this channel \cite{DeLuca:2020sae} or potentially being excluded by constraints from CMB anisotropies \cite{ser}. We will consider this as a  benchmark value throughout this work. However, depending on the strength of PBH accretion before reionization epoch, corrections to both the efficiency of the primordial binary merger rate \cite{DeLuca:2020sae} and the constraints on the PBH abundance may arise \cite{DeLuca:2020fpg}. If we take the DM local abundance to be of the order of the baryonic one, this would imply  a PBH local overdensity with respect to the DM in the GC of the order of ten.

\begin{table}
	\centering
	\begin{tabular}{c| c c c}
		\hline
		\hline
		GC & $N_\text{\tiny 10}$ & $N_\text{\tiny 60}$ & $N_\text{\tiny 80}$ \\
		\hline
		NGC 104   & 2294 & 2 & 2 \\     
		Pal 2     & 1628 & 1 & 1 \\     
		NGC 5139  & 1532 & 1 & 2 \\    
		NGC 6266  & 1639 & 1 & 1 \\
		NGC 6388  & 1439 & 1 & 1 \\    
		NGC 6441  & 2625 & 2 & 1 \\     
		Terzan 10 & 5484 & 4 & 3 \\     
		NGC 6715  & 6205 & 4 & 4 \\     
		NGC 7078  & 2270 & 2 & 2 \\
		\hline
		Terzan 10 & 2061 & 1 & 3 \\
		NGC 6715  & 1616 & 1 & 4 \\
		\hline
		\hline
	\end{tabular}
	\caption{Characteristic  maximum number of each BH species inside MWGCs we consider in this work. The first nine rows have been evaluated with $w_\text{\tiny kick}=50$ km/s and the last two with 100 km/s. We indicate only those WMGCs which contain a non-zero integral number of 60 $M_\odot$ ABHs and include in this table the value rounded to the nearest integer. For the last column we take $f_\text{\tiny 80}=10^{-4}$.}
	\label{table_BHs}
\end{table}

Finally, the segregation time of an object mass $m$ scales like $\tau_\text{\tiny seg}(m)\simeq ({\langle m\rangle /m})\tau_\text{\tiny relax}$, where $\tau_\text{\tiny relax}$ is the half mass relaxation time which is at most 1 Gyr for most GCs, see Ref.~\cite{2015ApJ...800....9M,OLeary:2005vqo,Fregeau:2001rp}. Thus, it only takes a few tens of Myr for the heaviest of the objects to sink into their segregation volumes. Also, the heavy progenitor stars that give rise to the ABH remnants have very short lifetimes, at most a few tens of Myr. For all these reasons we may assume as an initial condition that BHs have segregated towards the center and occupy their segregation volumes.

\subsection{Third-body hardening merger rate}

According to current understanding, isolated BH binaries can only merge through the emission of gravitational waves (GWs). The characteristic merger time is given by the Peter's timescale in terms of the initial semi-major axis (SMA) $a$, high initial eccentricity $e$ and BH masses $m_1$ and $m_2$ as \cite{peters1964}
\begin{align}\label{petergw}
T_\text{\tiny GW} &\simeq \frac{3}{85}\frac{c^5 a^4} {G^3 m_1 m_2 (m_1+m_2)} (1-e^2)^{7/2}\\&\simeq 8.5\times10^{11} \text{yr}\left({a\over\text{AU}}\right)^4\left({m_1\over80M_\odot}\right)^{-1}\left({m_2\over60M_\odot}\right)^{-1}\left({m_1+m_2\over140M_\odot}\right)^{-1}(1-e^2)^{7/2}\nonumber.
\end{align}
This can be much larger than the lifetime of the universe even for highly eccentric compact binaries, see Ref.~\cite{peters1964}. For instance, to gain a feeling of the timescales involved, for a BH-BH binary with equal masses  it takes about 
\begin{equation}
	T_\text{\tiny GW}(e=0.9)\simeq5.4\times10^{11} \, {\rm yr}\lp{a\over1\text{AU}}\rp^{4}\lp{m\over10M_{\odot}}\rp^{-3} 
\end{equation}
 for the BHs to coalesce due to GW emission alone, which is more than ten times the current age of the universe. Nevertheless, the situation is different when the same binary evolves inside a dense astrophysical environment, in the core of a stellar cluster for example. In particular, encounters of the binary with third bodies may in fact accelerate the merging process during the inspiral phase when the GW emission is still inefficient to drive the binary to merge \cite{Kritos:2020fjw,Sesana:2015haa}. The binary therefore can shrink fast in the initial phase of its evolution, when its SMA axis is still large enough for the interaction cross section to be high. 

According to the Heggie-Hills law, if the binary is tight enough, during the interactions of the binary with a single third body, some energy is transferred to the single and the binding energy of the binary increases, making it tighter. The opposite holds for soft binaries, which get softer~\cite{1975MNRAS.173..729H, 1980AJ.....85.1281H}. The SMA becomes smaller and smaller until the two members of the binary are so close that GW emission starts to dominate and takes over its evolution. This occurs at a SMA which we denote $a_\text{\tiny GW}$. It is important to note that during these flybys, the binary recoils and it is possible that the binary escapes and merges outside the dense environment~\cite{Samsing:2017plz, Samsing:2017xmd, Samsing:2019dtb, Samsing:2020qqd}. However, as long as the characteristic SMA required for ejection $a_\text{\tiny  ej}$ is smaller 
than  $a_\text{\tiny  GW}$, the binary merges within the cluster~\cite{Baibhav:2020xdf}, which is the case for the MWGCs and the BH binaries under consideration.

Consider a binary with mass components $m_1$ and $m_2$, with $m_1\ge m_2$. As this binary encounters single objects the outcome of the interaction could result in the increase or decrease of its binding energy, in the exchange of one of the components, or even ionization of the binary, see Refs.~\cite{1983ApJ...268..319H,1975MNRAS.173..729H}. To
account for interactions  for which the shrinking of the SMA is the most probable outcome, we need to make two further assumptions. First, we only account for the hardening process due to the binary interactions  with single objects of mass $m_3\ll m_2$. In our context these would be stars whose mass is at most $\sim M_\odot$ and a lot smaller than that of the BHs we examine. Second, we consider binaries with a SMA small enough to suppress the disruption probability of the binary. We define therefore hard binaries as those with a SMA smaller than \cite{Sesana:2006xw} 
\begin{align}
	a_\text{\tiny H}\equiv {G m_2 \over 4\sigma_*^2}\simeq 2.23\,  \text{AU}\lp{m_2 \over M_{\odot}}\rp\lp{\sigma_*\over10 \text{km/s}}\rp^{-2},
\label{smaHard}
\end{align}
for which the binary binding energy is greater than the kinetic energy of the cluster particles,
where $\sigma_*$ is the rms velocity of the ambient stars. 
Such a tight binary hardens at a constant rate, see for example Refs.~\cite{1975MNRAS.173..729H,Quinlan:1996vp,Sesana:2015haa,1980AJ.....85.1281H},  whose corresponding parameter $H$ is determined by numerical surveys  to lie in the range $(15\div 20)$~\cite{Quinlan:1996vp,Sesana:2015haa}. Here, throughout our calculations we use the value of $H=15$, however a different choice of $H$ in the prescribed interval does not change our results significantly.
The semi-analytic formula for the merger timescale of the binary with initial eccentricity $e$ via hardening with single third bodies is \cite{Kritos:2020fjw,Sesana:2015haa}
\begin{align}
	T_{1-2}^\text{\tiny 3rd}&=\left(\frac{5}{64}\right)^{1/5}\frac{c}{G^{7/5}}\left(\frac{\sigma_{3,1-2}}{\rho_3 H}\right)^{4/5}\left(m_1m_2(m_1+m_2)\right)^{-1/5}F(e)^{-1/5}\label{Tm_3rdbody} 
\nonumber	\\
	&\simeq 1.65 \ \text{Gyr} \lp{\sigma_{3,1-2}\over10 \text{km/s}}\rp^{4/5}\lp{\rho_3\over10^5M_\odot/\text{pc}^3}\rp^{-4/5}\lp{H\over15}\rp^{-4/5} \nonumber \\&\hspace{1.2cm}\times\lp{m_1\over80 M_\odot}\rp^{-1/5}\lp{m_2\over60M_\odot}\rp^{-1/5}\lp{m_1+m_2\over140M_\odot}\rp^{-1/5}F(e)^{-1/5}.
\end{align}
Here, $\rho_3$ is the local mass density of third single objects and $\sigma_{3,1-2}$ is the relative rms velocity between the binary and the single. Also, in Eq.~\eqref{Tm_3rdbody}, we use the auxiliary function defined in Ref.~\cite{peters1964} as $F(e)=(1-e^2)^{-7/2}\left(1+{73}e^2/{24}+{37}e^4/{96}\right)$. Given a time window $T$, at most equal to the cluster lifetime which corresponds to the formation timescale of a hard binary, the 3rd-body hardening merger rate per cluster is given by \cite{Kritos:2020fjw}
\begin{equation}
\label{eq: hardrate}
	\Gamma_\text{\tiny 3rd}=N {f_e(T_\text{\tiny cl}-T)\over T_\text{\tiny cl}-T}\Theta(T_\text{\tiny cl}-T),
\end{equation}
where $N$ is the number of available hard binaries of a specific type and $f_e$ is the fraction of those binaries which have an eccentricity larger that the critical value for which $T^\text{\tiny 3rd}_{1-2}=T_\text{\tiny cl}-T$.

In the next two sections we present two mechanisms that were considered for the calculation of the formation of a compact  pair by previous works and we estimate the corresponding merger rates. For example, recently Ref.~\cite{Nitz:2020bdb} have concentrated on the direct capture (DC) of an ABH  with a sub-solar mass PBH, while Ref.~\cite{Tsai:2020hpi} studied the direct capture of a neutron star with a PBH specifically focusing on the multi-messenger event GW170817~\cite{TheLIGOScientific:2017qsa}. 
Also, a different mechanism has been proposed in the literature where a binary may be induced by the encounter of three objects interacting within the same vicinity. The newly formed binary from this 3-body induced mechanism later undergoes 3rd-body hardening and subsequently merges. We refer to this binary formation mechanism as the 3-induced (3-ind.) channel. We repeat the calculations for the DC in Sec.~\ref{DC_section} and the 3-ind. channel in Sec.~\ref{3ind_section} in the case of GW190521 taking into account the MWGCs. We will conclude  that these mechanisms are not able provide the merger rate required to explain the mass gap event in question assuming the heavier BH, which is within the mass gap, to be primordial. This will bring us to Sec.~\ref{catalysis_section} in which we present a more efficient channel for PBH-ABH mergers.

\section{The direct-capture channel}
\label{DC_section}
\begin{figure}[t]
	\centering
	\includegraphics[width=.8\textwidth]{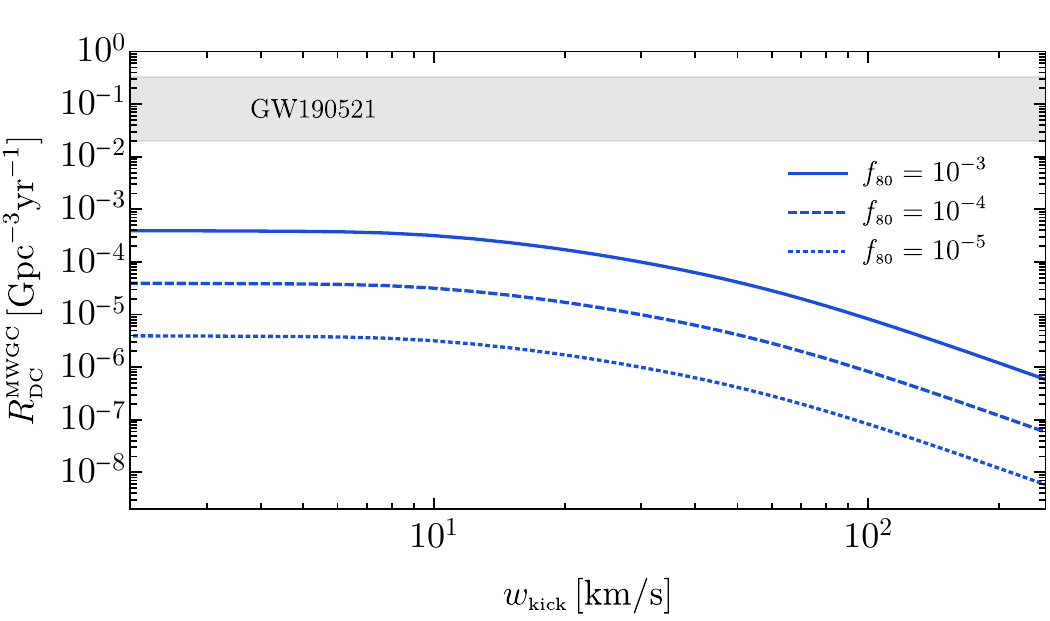}
	\caption{The DC merger rate density as a function of the natal kick velocity parameter for three values of $f_\text{\tiny 80}$. The results scale proportional to the number density of GCs normalized to $2.4 \,\text{Mpc}^{-3}$. The horizontal grey strip defines the LIGO/Virgo empirical merger rate density of GW190521.}
	\label{DC_fig}
\end{figure}

\noindent
The DC channel is a relativistic mechanism through which two objects encounter themselves with such a small pericenter so that GW dissipative effects subtract an amount of energy from the system leaving behind a bound pair. In particular, if two objects with masses $m_1$ and $m_2$ interact with a pericenter smaller than a maximum value given by  \cite{quin,mouri}
\begin{align}
	r_p^\text{\tiny DC}&=\lp{85\pi\sqrt{2}G^{7/2}m_1m_2(m_1+m_2)^{3/2}\over 12c^5\sigma_{1,2}^2}\rp^{2/7} \label{rpDC}
	\nonumber \\
	&\simeq 9\times10^{-4}\, \text{AU}\lp{m_1\over60M_{\odot}}\rp^{2/7}\lp{m_2\over80M_{\odot}}\rp^{2/7}\lp{m_1+m_2\over140M_{\odot}}\rp^{3/7}\lp{\sigma_{1,2}\over10\text{km/s}}\rp^{-4/7}, 
\end{align}
then a binary forms. Notice that for the conventional values of masses and relative velocity considered in this paper,  Eq.~\eqref{rpDC} gives a maximum pericenter distance which is very small compared to the sizes of most stellar systems. Also, a binary formed through this capture process is highly eccentric \cite{Cholis:2016kqi,OLeary:2008myb,Samsing:2019dtb} and thus merger almost instantly. Applying the Peter's GW formula \eqref{petergw} to a pair with a SMA of $a=r_p^\text{\tiny DC}/2$, the merger timescale in the highly eccentric limit is given by
\begin{equation}
T_\text{\tiny GW}(e)\simeq0.05\ \text{yr} \lp{a\over5\times10^{-4}\text{AU}}\rp^4(1-e^2)^{7/2}.
\end{equation}
 Therefore, once a pair of objects capture themselves, the binary merges  quickly before the encounter with a third body which might perturb the orbit. By defining $n_1$ and $n_2$ as the number densities of each population and $\sigma_{1,2}$ as their rms relative velocity, then we can write the differential DC merger rate per unit volume and per cluster as \cite{quin,mouri}
\begin{equation}
	\gamma_\text{\tiny DC}\simeq 
	17\,   n_1n_2{G^2\over c^{10/7}}{(m_1+m_2)^{10/7}(m_1m_2)^{2/7}\over\sigma_{1,2}^{11/7}}.
\label{diff_DC}
\end{equation}
This formula was obtained performing a statistical  average  assuming  a Maxwellian distribution for  the  velocities.
To evaluate the total DC merger rate per cluster one should integrate Eq.~\eqref{diff_DC} over the volume. However, since in our case we assume that the BH species are uniformly distributed throughout their segregation volumes, it suffices to multiply the differential merger rate by the minimum segregation volume between the components $m_1$ and $m_2$ as
\be
\Gamma_\text{\tiny DC} = {4\pi\over3} \text{min}\llp R_{\text{\tiny seg}}^3(m_1),R_{\text{\tiny seg}}^3(m_2)\rrp\gamma_\text{\tiny DC}.
\ee
 This is because a capture event between BHs can only take place inside the intersection of their regions where the two populations are distributed. Finally, to account for the merger rate density, we simply average over the MWGCs\footnote{We generically denote the average of a rate $\Gamma$ over the MWGCs taken from the Harris catalog \cite{harris2010new} as $\langle \Gamma \rangle_\text{\tiny  MWGC} \equiv \sum_{j=1}^{N_\text{\tiny MWGC}} \Gamma_j/N_\text{\tiny MWGC}$.
}
 and then multiply by their number density $n_\text{\tiny GC}$ in the local universe to get
\begin{equation}
	R^\text{\tiny MWGC}_\text{\tiny DC}=n_\text{\tiny GC} \langle \Gamma_\text{\tiny DC} \rangle_\text{\tiny MWGC}.
	\label{R_DC}
\end{equation}
In Fig.~\ref{DC_fig} we evaluate the DC merger rate density of a $60M_\odot$ ABH and an $80M_\odot$ PBH 
for Milky way type globular clusters by averaging over the $N_\text{\tiny MWGC}=140$ environments in the MWGC catalog \cite{harris2010new}.
 We vary the results depending on  $w_\text{\tiny kick}$ and three values of $f_\text{\tiny 80}$ to show the behavior of the final result as a function of the parameters.

The dependence of the DC merger rate with the PBH abundance scales as
\begin{align}
	R_\text{\tiny DC}^\text{\tiny MWGC}& \simeq
	3 \cdot 10^{-6} \, \text{Gpc}^{-3}\text{yr}^{-1}
	\lp \frac{f_\text{\tiny 80}}{10^{-4}} \rp
\lp n_\text{\tiny GC}\over2.4 \, \text{Mpc}^{-3}\rp,
\quad
 \text{for}
 \quad w_\text{\tiny kick}=50\,\text{km/s}.
\end{align}
We observe that for a modest choice of $f_\text{\tiny 80}$ at a value below $10^{-4}$, the DC channel predicts much lower merger rates compared to the observed rate for the mass gap event $R\simeq 0.1{\rm  Gpc^{-3} yr^{-1}}$. The reason is that, as we can see in Eq.~\eqref{rpDC}, the maximum  pericenter required for the DC mechanism is much smaller than  $\sim 1$ AU. This leads to a tiny DC cross section between a $60M_\odot$ ABH and an $80M_\odot$ PBH even if they  segregate in the core of dense GC environments.

\section{The three-body induced channel}
\label{3ind_section}
\noindent
Unlike the DC channel, the 3-ind.  one  is a non-dissipative mechanism through which binaries can form during a three body encounter. When two objects approach from infinity and interact with a pericenter that is larger than the DC pericenter $r_p^\text{\tiny DC}$ (Eq.~\eqref{rpDC}) then GW emission is not strong enough to induce a binary. 
According to energy conservation considerations, the two objects will therefore  escape back to infinity rather than capturing themselves. However, if a third object also participates
in the encounter then a 3-body interaction will occur. For the rate of 3-body encounters, one has to multiply the interaction rate between two objects with the probability of the third object to be inside the same vicinity. The resulting differential rate for a third object of mass $m_3$ to participate in a 3-body encounter within a region of size $b$, where two other objects with masses $m_1$ and $m_2$ already interact, was calculated in \cite{Ivanova:2005mi} to be
\begin{equation}
	\gamma_\text{\tiny 3-body}= n_1n_2n_3\sigma_{1,2}\pi b^5\left[\sqrt{8\over3\pi}+\sqrt{24\over\pi}{G(m_1+m_2)\over\sigma_{1,2}^2b}\right]\lp 1+\sigma_{3,1-2}\sqrt{4b\over3\pi G(m_1+m_2)}\rp,
\label{3-body_eq}
\end{equation}
where an average over the relative velocities has been performed.
Here, $\sigma_{1,2}$ is the rms relative velocity between 1 and 2 and $\sigma_{3,1-2}$ the rms relative velocity of 3 with respect to the center of mass of the 1-2 system.
As in the previous section, we will restrict ourselves to the case in which $m_3 \ll m_2, m_1$. Also,
 to obtain the total 3-body formation rate we integrate over volume. In our context we simply multiply Eq.~\eqref{3-body_eq} by the minimum segregation volume of the objects with  masses $m_1$ and $m_2$
 \be
 \Gamma_\text{\tiny 3-body} = {4\pi\over3} \text{min}\llp R_{\text{\tiny seg}}^3(m_1),R_{\text{\tiny seg}}^3(m_2)\rrp\gamma_\text{\tiny 3-body}.
 \label{3-body_eq2}
 \ee
Notice that Eq.~\eqref{3-body_eq2} does not describe the binary formation rate, because not all of the 3-body encounters efficiently produce binaries. That is, Eq.~\eqref{3-body_eq2} should be multiplied by an efficiency factor $\eta\in(0,1)$ that determines the fraction of 3-body encounters that result in the formation of a binary. This fraction would depend on the velocities and the masses of the objects participating in the 3-body encounter.
Numerical simulations have been performed in the past, for instance in Refs.~\cite{1976A&A....53..259A,1971SvA....15..411A,2010ApJ...717..948I} to investigate the binary formation probability by 3-body encounters. In particular,
Refs.~\cite{1976A&A....53..259A,1971SvA....15..411A} found that for equal mass objects, 
the binary formation rate is high and even closer to $\eta=1$ when the 3 bodies encounter in a very small region of size $b$ corresponding to the SMA of an induced hard binary. 
In the more general case, the formula in Eq.~\eqref{3-body_eq2} provides only an upper limit for number of 3-body induced binaries per unit time.
If a binary is induced in a 3-body encounter then we expect its maximum SMA to be of order $b/2$.
In order to maximize the estimate for the  binary formation rate through this channel but to also ensure that the newly formed binary survives after its formation from this channel, we take $b=2a_\text{\tiny H}$. 
Once these binaries have formed inside a dense environment, then they evolve via the 3rd-body channel and harden until they merge. 

This channel requires both the $60 M_\odot$ and the $80 M_\odot$ BHs to be single inside the cluster. However, one needs to account for the  depletion of
the population of single BHs on a timescale of $T_\text{\tiny 3-ind}=\min\left(\Gamma_\text{\tiny 3-body}^{-1},\Gamma_\text{\tiny A}^{-1}N_\text{\tiny 60-10},\Gamma_\text{\tiny B}^{-1}N_\text{\tiny 80-10}\right)$  as
these compact objects tend to obtain a companion and form a binary. Usually in the early stages of the cluster they obtain the more abundant $10 M_\odot$ ABH.
In this timescale, $\Gamma_\text{\tiny 3-body}$, $\Gamma_\text{\tiny A}$ and $\Gamma_\text{\tiny B}$
stand for the competing interaction rates leading to the formation of 80-60, 60-10 and 80-10 from a 10-10 progenitor, respectively
(see Appendix.~\ref{algorithm_appendix} for more details).
 The corresponding 3-ind. differential merger rate per environment will be 
\begin{equation}
\Gamma_\text{\tiny 3-ind} = \eta \Gamma_\text{\tiny 3-body} \frac{T_\text{\tiny 3-ind}}{T_\text{\tiny cl}-T_\text{\tiny 3-ind}}  f_e\lp T_\text{\tiny cl}-T_\text{\tiny 3-ind} \rp  \Theta\lp T_\text{\tiny cl}-\Gamma_\text{\tiny 3-body}^{-1}\rp,
\label{3-ind.dens}
\end{equation}
where we stress the inclusion of the efficiency factor $\eta$ as discussed above.
Here, the theta function is included to ensure that an ABH-PBH pair have formed on average on a timescale shorter than the lifetime of the cluster.

\begin{figure}[t]
	\centering
	\includegraphics[width=.8\textwidth]{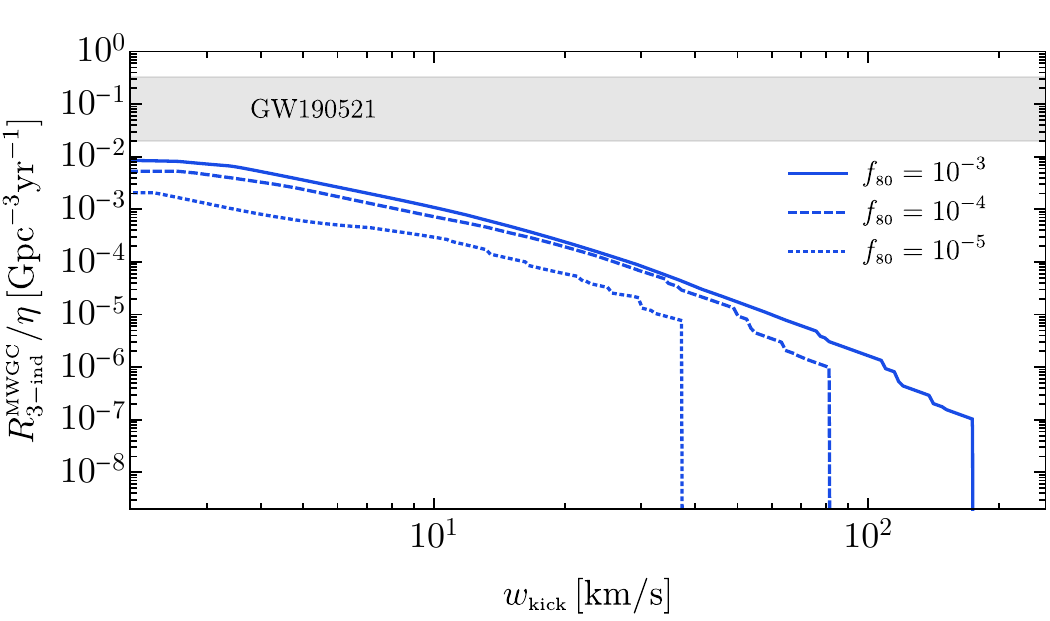}
	\caption{The 3-ind. merger rate density as a function of the natal kick velocity parameter for three values of $f_\text{\tiny 80}$. The results scale proportional to the number density of GCs normalized to $2.4 \, \text{Mpc}^{-3}$. We normalize the vertical axis to the probability for binary formation $\eta\in(0,1)$ as defined in the text. The hard cutoff is due to the theta function in Eq.~\eqref{3-ind.dens} and the horizontal grey strip defines the LIGO/Virgo empirical merger rate density of GW190521.}
	\label{3ind_fig}
\end{figure}

In our context, we evaluate the 3-ind. merger rate of a $60M_\odot$ ABH with an $80M_\odot$ PBH forming a pair through 3-body encounters. We refer to such a binary as an 80-60 pair. To maximize the probability an ABH and a PBH interact with a third object in the cores of MWGCs, we take as the third object to be a star since those are the most relevant component in the environment we consider.
The final  merger rate density in the local universe from this channel is given by
\be
	R^\text{\tiny MWGC}_\text{\tiny 3-ind}=n_\text{\tiny GC}  \langle \Gamma_\text{\tiny 3-ind} \rangle_\text{\tiny MWGC}.
\ee
We present our findings in Fig.~\ref{3ind_fig}. We find that this channel can be active only in a few systems; in particular for $f_\text{\tiny 80}=10^{-4}$ only the following MWGCs contribute with a non-negligible 3-ind. merger rate: Pal 1, NGC 6256, Terzan 2, HP 1, Terzan 1, NGC 6397, Terzan 9, NGC 6544, NGC 6558, and Pal 12. These are the most compact MWGCs with the highest central stellar densities.
As the term $f_\text{\tiny 80}$ is lowered then even fewer systems participate non-trivially in the average value. 

We conclude that this channel requires that, for a value of the fraction $f_\text{\tiny 80}$, the natal kick $w_\text{\tiny kick}$ should be below some threshold value in order for the ABH-PBH pair to form in the core of a few GC systems within their lifetime. Above that critical value the rate drops to zero due to the hard cutoff we imposed in Eq.~\eqref{3-ind.dens}.
For considerations similar to the ones described for the DC channel,  $f_\text{\tiny PBH}$ is constrained to be below ${\cal O}(10^{-5})$ and for reasonable values for the fraction $f_\text{\tiny 80}$ this channel appears ineffective in explaining a GW190521-like merger rate. Even though we report that this channel predicts higher rates than the DC case, we stress again that our results are only an upper limit and the fiducial value may not be closer to the empirical interval expected for GW190521. 
Furthermore, it seems that for $f_\text{\tiny 80}<10^{-5}$ the merger rate drops above small values of the $w_\text{\tiny kick}$. This indicates that if ABHs tend to obtain a natal kick larger than the point where the 3-ind. rate drops, then we need to look for other mechanisms in order to explain events like GW190521.

\section{The catalysis mechanism}
\label{catalysis_section}
\noindent
Given the failure of the previously discussed two channels, we focus in this section on another possibility, which we call ``catalysis" mechanism and, to the best of our knowledge, has never been investigated so far.

We focus on the formation of high mass ABH-PBH binaries through a process of successive exchanges during binary-single and binary-binary interactions. We remind the reader that for illustrative purposes, in this work we consider a model with  a PBH population with a monochromatic mass spectrum at $m_\text{\tiny 80}=80M_\odot$ and an ABH population with a double peak mass spectrum with a small mass component at $m_\text{\tiny 10}=10M_\odot$ and a high mass component at $m_\text{\tiny 60}=60M_\odot$. The relative sizes and abundances of these populations are given by Eqs.~\eqref{N_80},~\eqref{N_10} and ~\eqref{N_60}. As an initial condition we consider a population of seed binaries with two $10M_\odot$ ABHs which we denote as 10-10 pairs. We denote by $f_\text{\tiny bin}$ the fraction of $10M_\odot$ ABHs that participate in such binaries. Furthermore, only a fraction $f_\text{\tiny hard}$ of these pairs will be hard. Therefore, the initial number of hard 10-10 binaries inside a cluster will be $N_\text{\tiny 10-10}^\text{\tiny (1)}={1\over2}f_\text{\tiny hard}f_\text{\tiny bin}N_\text{\tiny 10}$ (see also Eq.~(11) in \cite{Kritos:2020fjw}). Also, initially we take the $m_\text{\tiny 60}$ and $m_\text{\tiny 80}$ to be single objects inside each cluster with initial population sizes denoted by $N_\text{\tiny 60}^\text{\tiny (1)}$ and $N_\text{\tiny 80}^\text{\tiny (1)}$ respectively. The purpose of this model is to test the efficiency of successive exchanges in forming high mass binaries in MWGCs which then merge and produce GW signals like the mass gap event GW190521.

Inside the dense cores of GCs, these 10-10 pairs interact with heavier BHs. In our model massive species will be $60M_\odot$ ABHs and $80M_\odot$ PBHs. Such 60-(10-10) and 80-(10-10) binary-single interactions can occur when a 10-10 pair passes nearby the center of the cluster and finds itself inside the segregated volume of massive objects, either the 60$M_\odot$ ABH or the 80$M_\odot$ PBH. During this time, a member of the 10-10 pair can be swapped for the $60 M_\odot$ (or $80 M_\odot$) object forming a 60-10 (80-10) binary respectively. Subsequently,  (60-10)-80, (80-10)-60 binary-single and (60-10)-(80-10) binary-binary encounters can form an 80-60 pair. 
All in all, there are in total six binary species considered here: 10-10, 60-10, 80-10, 60-60, 80-80 and 80-60.

\begin{figure}[t]
	\centering
	\includegraphics[width=0.9\textwidth]{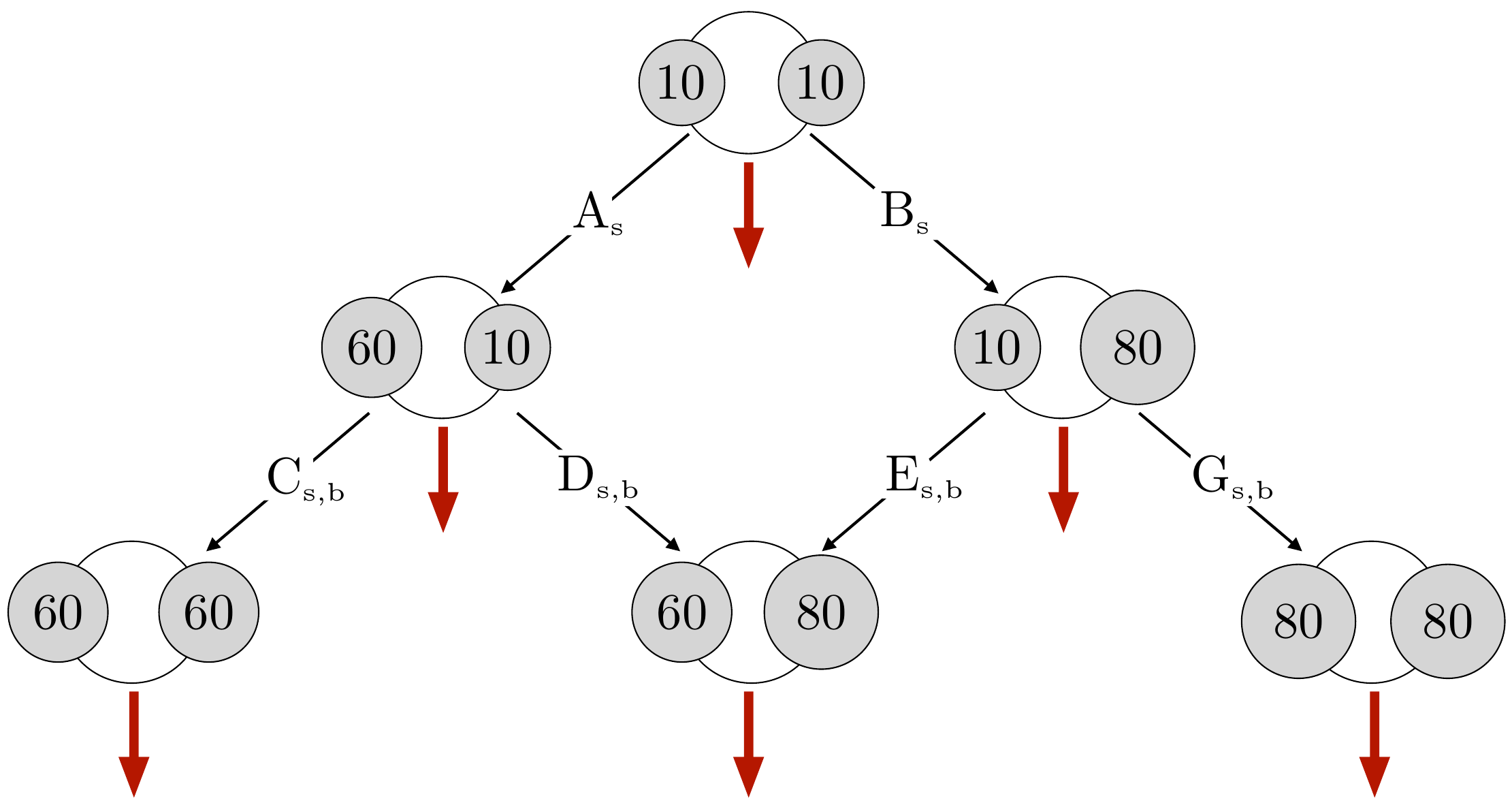}
	\caption{Representation of our model as discussed in Sec.~\ref{catalysis_section}. The value shown in each member of the binary populations in this figure indicates its mass in units of $M_\odot$. Each thin black arrow shows the direction a binary can evolve into, they are labeled by a capital Roman letter and the subscript denotes whether that process can proceed via a binary-single (``s'') and/or a binary-binary (``b'') interaction. The red arrows denote the merger of a given population.}
	\label{sketch}
\end{figure}

In Fig.~\ref{sketch} we present a graph of the possible pathways examined in this work that can lead to the various populations of binary BHs through a cascade of exchange processes. All of these processes, except A and B, can be implemented by both binary-single and binary-binary exchanges. So in principle all processes, but the A and B which can only occur as binary-single exchanges, can be expressed as the sum of two components. For instance, if ``s'' refers to binary-single and ``b'' to the binary-binary case respectively, we write $\Gamma_\text{\tiny D}\equiv\Gamma_{\text{\tiny D}_\text{\tiny s,b}}=\Gamma_{\text{\tiny D}_\text{s}}+\Gamma_{\text{\tiny D}_\text{\tiny b}}$ for the exchange rate of 60-10$\to$80-60 corresponding to the arrow named ``D$_\text{\tiny s,b}$'' in Fig.~\ref{sketch}, or simply process ``D''. Note that in this particular example, the binary-single interaction will be of the form (60-10)-80, while the binary-binary interaction will be between 60-10 and 80-10. Moreover, we are not interested in the fate of the remnant, whether they are ejected from the cluster or 
retained and  participate in second generation mergers.
In practice we are considering only first generation mergers.
 We describe in detail the semi-analytic calculations we implement in Appendices~\ref{exchanges_appendix} and~\ref{algorithm_appendix}.

\begin{figure}[t]
		\centering
		\includegraphics[width=1\textwidth]{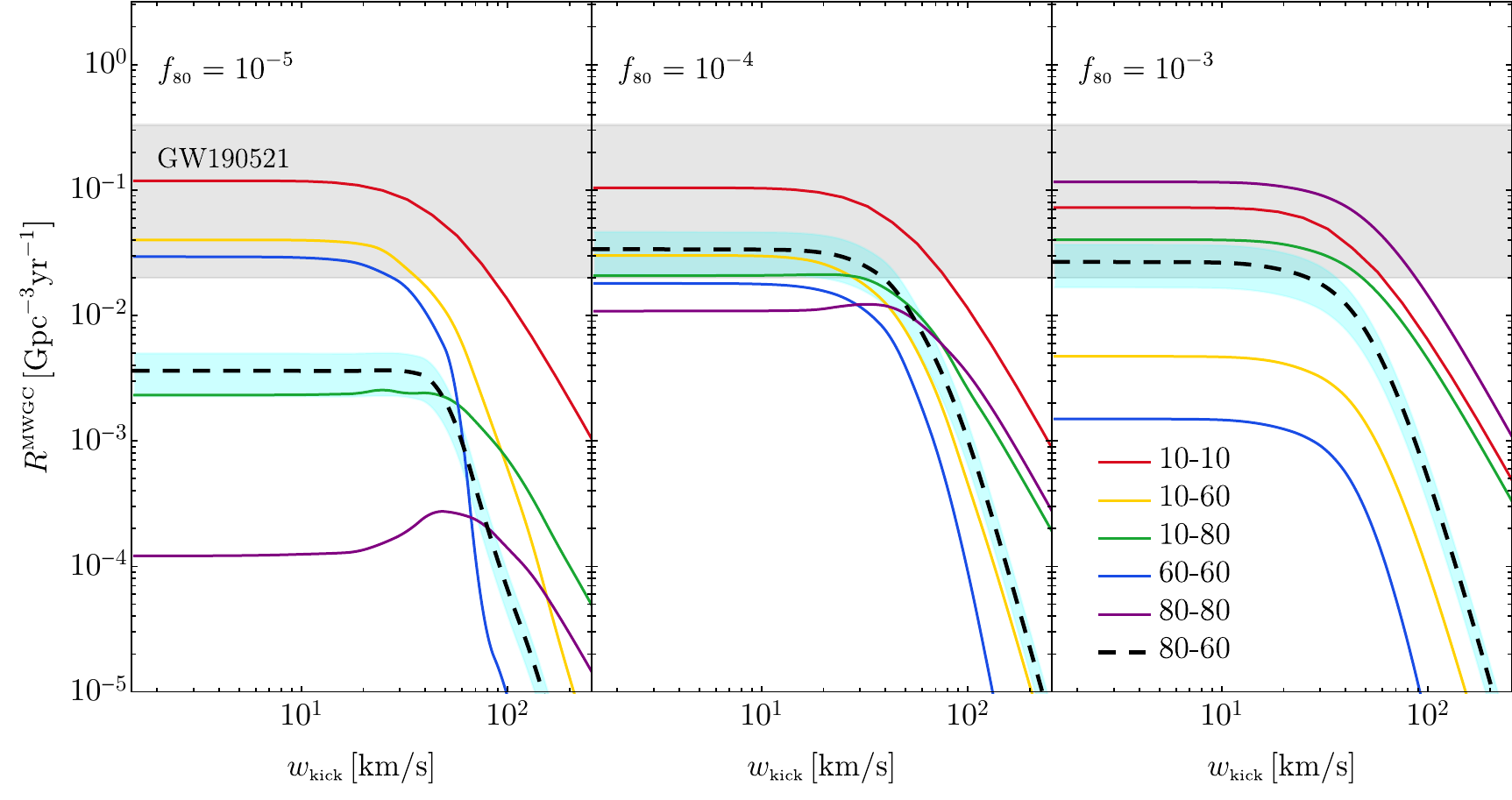}
	\caption{The 3rd-body merger rate densities of the various binary populations we consider in the catalysis channel, for different values of $f_\text{\tiny 80}=10^{-5},10^{-4},10^{-3}$, respectively.
	The cyan band represents the uncertainty associated to $n_\text{\tiny GC} = 2.4^{+0.9}_{-0.9}\,\text{Mpc}^{-3}$ and the transparent grey band corresponds to the merger rate density interval inferred for GW190521, $0.13^{+0.30}_{-0.11}\text{Gpc}^{-3}\text{yr}^{-1}$ ($90\%$ credibility) by the LIGO/Virgo collaboration. 
	}
	\label{Final_results}
\end{figure}

In Fig.~\ref{Final_results} we show the results that we obtain from running the catalysis algorithm. 
The initial abundance of 10-10 binaries is set to a value of  $f_\text{\tiny hard}f_\text{\tiny bin} \approx 0.3\%$,  as motivated by recent numerical surveys in Refs.~\cite{Park:2017zgj, Chatterjee:2016hxc}. Indeed, as shown in Ref.~\cite{Chatterjee:2016hxc}, the number of BH binaries is found to be almost constant in time against the subsequent evaporation of the clusters. In particular, 
it was found that a cluster with a large initial number of BHs will experience a constant dynamical production and subsequent destruction of binaries, given that those can be ionised by dynamical interactions with single BHs in the environment. As time goes, the evaporation of the clusters leads to a smaller number of isolated BHs, with the consequent creation of a smaller number of binaries, which are however subject to fewer interactions with single BHs. These combined effects lead to an almost constant number of retained BH binaries in the clusters. 
	Given that in our work we are not following the time evolution of the globular clusters, which would need a dedicated numerical simulation, we have kept into account its impact in an effective way by fixing the fraction $f_\text{\tiny hard}f_\text{\tiny bin}$ of hard retained BH binaries  such that their number is compatible with the one obtained in the numerical simulation performed in Ref.~\cite{Chatterjee:2016hxc}. Let us stress here that the heavier BH species, like the 60$M_\odot$ and 80$M_\odot$ BHs, are more easily retained due to their large mass with respect to the dominant population of 10$M_\odot$ BHs and therefore less 
	subject to the evaporation phenomenon.

Below some threshold value of $w_\text{\tiny kick}$ the curves in Fig.~\ref{Final_results} reach a plateau since $f_\text{\tiny ret}$ saturates to $100\%$. The drop in the rates for higher values of $w_\text{\tiny kick}$ is due to the fact that fewer ABHs are retained when the natal kick is high. For high values of $w_\text{\tiny kick}$ the drop-off is a power law of the form $R^\text{\tiny MWGC}\propto w_\text{\tiny kick}^q$, where $q$ can be inferred for each merger rate by measuring its slope. In particular, we find that $q_1\simeq-2.89$ for 10-10, 80-10 and 80-80 mergers, $q_2\simeq-5.94$ for 60-10 and 80-60 and $q_3\simeq-8.94$ for the 60-60 mergers. 
The hierarchy  $q_3<q_2<q_1$ is expected as the number of interactions involving ABHs increases along a particular merger chain, see Fig.~\ref{sketch}, and therefore the corresponding merger rates are suppressed by higher powers of the retention fraction.

Focusing on $f_\text{\tiny 80}=10^{-4}$, we can find a subset in the parameter space in which our estimate lies inside the LIGO/Virgo interval of $0.13^{+0.30}_{-0.11}\text{Gpc}^{-3}\text{yr}^{-1}$ for a GW190521-like binary. 
As already highlighted, the merger rate is impacted by the retention of ABHs. 
One can notice that the 80-60 merger rate in Fig.~\ref{Final_results} falls below the observed band when $w_\text{\tiny kick}>50$ km/s (which would correspond to a retention fraction of $f_\text{\tiny ret}>20\%$ assuming an average escape velocity of $V_\text{\tiny esc}=50$ km/s).  For $f_\text{\tiny 80}=10^{-4}$, we observe that the 80-60 merger rate stands near the bottom of this experimental interval, and therefore 
for smaller abundance of PBHs inside GCs the mixed scenario becomes subdominant as are all the rates of binary mergers involving at least one PBH.

In addition, for large values of $w_\text{\tiny kick}$, the 80-80 merger rate systematically overcomes both the 80-60 and 60-60 rate, as can be seen from Fig.~\ref{Final_results}. 
This is because, as $w_\text{\tiny kick}$ is increased, the processes B and G dominate over the others as fewer 60s are retained in each cluster. 
This is the origin of the slight increase in the 80-80 merger rate around $w_\text{\tiny kick} \simeq {\cal O}(10^2)$ km/s. 
Since a smaller fraction of $60 M_\odot$ ABHs is retained, the numbers of 60-10 binaries and single 60 are also diminished and the most probable fate of an 80-10 binary is to interact with another 80-10 and likely give an 80-80 pair which then merges.

We can give a rough estimate of the 80-60 merger rate density for values of $w_\text{\tiny kick}<30$ km/s where the rate is saturated. Independently of $w_\text{\tiny kick}$, this scales linearly with the PBH abundance and can be written as
\begin{align}
	R^\text{\tiny MWGC}_\text{80-60}\sim 0.03 \, \text{Gpc}^{-3}\text{yr}^{-1}\lp{f_\text{\tiny 80}\over10^{-4}}\rp\lp{n_\text{\tiny GC}\over2.4\,\text{Mpc}^{-3}}\rp.
\label{80+60_estimate}
\end{align}
 Note that this approximation holds true only for $f_\text{\tiny 80}<10^{-4}$ and should not be extrapolated for larger values of PBH abundance. 

The value of $f_{80}=10^{-4}$ defines a theoretical maximum for another reason as well. In particular, as we can see from the right-most panel of Fig.~\ref{Final_results} for which $f_{80}=10^{-3}$, the merger rates of 10-80 and 80-80 overcome the 80-60 rate. This is because the abundance of PBHs becomes so large that the processes B and G dominate over all cases. This is problematic because the predicted merger rate density for 80-80 would then be $\sim0.1 \, \text{Gpc}^{-3}\text{yr}^{-1}$ which is above the mean empirical value inferred by LIGO/Virgo for the similar GW190521-like events. We also observe that the 80-60 merger rate decreases as we consider environments with a larger abundance of PBHs in the cluster as the majority of 10-10 binaries will encounter more frequently single 80s than 60s, reducing the number of 10-60 and 80-60 binaries, with a consequent decrease of their corresponding merger rate.
 As a last remark, we note that the 80-80 merger rate density scales like $\propto f_{80}^2$ since there are two PBHs involved in the formation of an 80-80 pair and the $f_{80}$ factor enters twice.

\section{Conclusions}
\label{conclusions_section}
\noindent
Motivated by the existence of a mass gap in the ABH mass spectrum, in this work we considered a population of PBHs with masses narrowly distributed around $80 \, M_\odot$. The PBHs can give rise to binaries with astrophysical objects in dense environment through dynamical formation channels.

We probed three different mechanisms through which an 80-60 PBH-ABH binary can form and merge: the direct capture events, the three body induced binaries (both already proposed in the literature) and  a novel mechanism, the ``catalysis'' channel,  in which the assembly of an 80-60 pair develops under a cascade process of binary-single and binary-binary exchanges starting from lower mass binaries.\footnote{A parallel numerical investigation of a catalysis-type mechanism in the context of low mass gap object-BH mergers motivated by GW190814 is also under study and it  will be presented elsewhere \cite{KCb}.} We found that the most dominant channel out of the ones discussed in this work is the ``catalysis'' channel. This happens because exchange episodes occur frequently inside the cores of dense stellar clusters and binary-single hardening is also efficient in such high density environments. The smallness of the impact parameter required for two compact objects to capture themselves and the rareness of three body encounters render the DC and 3-ind. merger rates too small to give an observable contribution. Therefore, the fastest way to form a high mass astro-primordial BH binary is via exchange episodes.

To summarize our findings, we have demonstrated that astro-primordial BH mergers similar to the mass gap event GW190521 should form inside of Milky Way type globular clusters and provide merger rates which may be consistent with current ground-based experiments. 
This is achieved if PBHs cluster efficiently in such stellar environments with an overdensity which is at least a few and the retention fraction of ABHs is non-negligible. For instance, for a high retention fraction of about $>20\%$ on average and a PBH effective abundance factor of $f_\text{\tiny 80}=10^{-4}$ per system,  the merger rate density of 80-60 astro-primordial BH binaries is about $0.03 \,\text{Gpc}^{-3}\,\text{yr}^{-1}$, which  falls inside the inferred experimental rate interval at $90\%$ credibility for a GW190521-like mass gap event. We further note that our results contain only the contribution from GCs and that the final rate might be even larger with the inclusion of all possible collisional stellar environments. Some of these are active galactic nuclei disks \cite{Samsing:2020tda,Tagawa:2020jnc}, nuclear star cluster at the center of galaxies \cite{Miller:2008yw,Fragione:2020han} and young stellar clusters \cite{PortegiesZwart:2004ggg}. 
We leave a detailed analysis of those environments for future work as well as generalising our findings to  extended  mass functions.  
In any case, since we demonstrated that the natal kick of the ABHs affects our estimations, we expect astro-primordial mergers to most likely occur inside of massive stellar systems with a large escape velocity.

Moreover, from our analysis one can make a prediction about the merger rates of other types of binaries with higher mass ratios. We considered 80-10 and 60-10 binaries, whose merger rate is found to be very similar to the one expected for the 80-60. The 80-10, which would also be a mixed astro-primordial BH merger, tends to dominate over the 80-60 for low values of the retention fraction. 
Nevertheless, due to the higher mass hierarchy, the ground-base GW detectors are less sensitive to such events and the detection rate would be reduced  by a factor ${\cal O}(0.05)$ with respect to the intrinsic merger rates shown in Fig.~\ref{Final_results}, 
 see appendix C for few details on the selection bias at LIGO/Virgo experiments.

 Let us also notice that, in addition to the mixed 80-60 binary, the scenario investigated in this work is also predicting a comparable merger rate for the 60-60 ABH binaries. This would be compatible with the few massive events in the GWTC-2 catalog with primary masses of the order of $60 M_\odot$~\cite{Abbott:2020niy}.

	\begin{center}
		{\bf  Acknowledgments}
	\end{center}
	\noindent
	We thank I. Cholis and P. Pani for useful feedback on the draft and for interesting discussions.
	We also thank the anonymous referee for constructive comments.
	V.DL., G.F. and A.R. are supported by the Swiss National Science Foundation 
	(SNSF), project {\sl The Non-Gaussian Universe and Cosmological Symmetries}, project number: 200020-178787.

	\appendix
	
	\section{Appendix: Exchange Interactions}
	\label{exchanges_appendix}
	\renewcommand{\theequation}{A.\arabic{equation}}
In this appendix we review the formalism used in Section \ref{catalysis_section} to describe binary-single and binary-binary interactions, which are involved in  the ``catalysis'' mechanism.

\subsection{Binary-single exchanges}
	\noindent
	Interactions between a binary and a single object may result in the exchange of one of the components of the binary for that single object. In particular, if the intruder is more massive than one of the binary members, then the probability for an exchange is high \cite{1980AJ.....85.1281H}. Consider a hard pair with SMA $a$ and mass components $m_1$ and $m_2$ which interact with a single mass $m_3$, then the cross section for the exchange of $m_1$ with $m_3$ can be written using the semi-analytic formula proposed in Ref.~\cite{H} as
	\begin{equation}
	\Sigma_\text{\tiny ex}=\frac{GM_{123}a}{v_{3,1-2}^2}\times f(m_1,m_2,m_3),
	\label{ExCrossSect_heggie}
	\end{equation}
	where $v_{3,1-2}$ is the relative velocity between the binary 1-2 and the single 3.
	The mass dependent form factor $f$ multiplying the gravitational focusing term is given by
	\begin{align}
	f(m_1,m_2,m_3)&\equiv{ m_3^{7/2}M_{32}^{1/6} \over M_{12}^{1/3}M_{13}^{5/2}M_{123}^{5/6}}
	\times\exp\Big(3.70+7.49x-1.89y-15.49x^2-2.93xy-2.92y^2\nn\\&\hspace{4.5cm}+3.07x^3+13.15x^2y-5.23xy^2+3.12y^3\Big),
	\end{align}
	where $x={m_1}/{M_{12}}$, $y={m_3}/{M_{123}}$ and $M_{ij(k)}=m_i+m_j (+ m_k)$ ($i,j,k=1,2,3$).
	
	The differential exchange rate per unit volume for binary-single interactions is given by multiplying the cross section in Eq.~\eqref{ExCrossSect_heggie} with the relative velocity $v_{3,1-2}$ and  both the number densities of the binary $n_{12}$ and the single $n_3$. By performing an average with respect to velocity over the Maxwellian distribution, one finally gets
	\begin{equation}
	\gamma_\text{\tiny ex}=\sqrt{6\over\pi}n_{12}n_3{GM_{123}a\over\sigma_{3,1-2}}\times f(m_1,m_2,m_3),
	\label{diff_exchange_rate}
	\end{equation}
	with $\sigma_{3,1-2}$ the three dimensional rms relative velocity between the 1-2 pair and the 3 body at infinity. The total exchange rate inside a uniform environment is finally obtained by multiplying Eq.~\eqref{diff_exchange_rate} with the minimum of the segregation volumes of the 1-2 and 3 systems as
	\begin{equation}
	\Gamma_\text{\tiny ex}=\frac{4\pi}{3}\min\llp R_\text{\tiny seg}^3(M_{12}),R_\text{\tiny seg}^3(m_{3})\rrp \gamma_\text{\tiny ex}.
	\end{equation}
	In the post-exchange state, one can deduce the SMA $a'$  of the new  3-2 binary by implementing conservation of energy and momentum and integrating over the relative angles in the exchange, finally getting
	\begin{align}
	a'=\left\{{m_1\over m_3}{1\over a}+{1\over Gm_3m_2M_{123}}\left({64\over\pi^4}m_1M_{32}\left[\left({m_3\sigma_3\over M_{32}}\right)^2+\left({M_{12}\sigma_{1-2}\over m_1}\right)^2\right]-m_3M_{12}(\sigma_3^2+\sigma_{1-2}^2)\right)\right\}^{-1}
	\label{postSMA}
	\end{align}
in terms of the SMA $a$ of the old pair 1-2. 

The eccentricity of the initial binary is expected to follow the thermal distribution $f(e) \d e=2e \, \d e$ with mean value $ \langle e \rangle = 2/3$, which will get shifted to higher values once the exchange has taken place. However, as the binary continues interacting with other compact objects in the GC, we expect the distribution to be consequently modified after exchange process.
Therefore, in order to give a conservative estimate for the rate, we will assume in our computation that also the eccentricity of the final binary follows the thermal distribution.

\subsection{Binary-binary exchanges}
\noindent
	Binary-binary encounters can also result in exchange episodes and formation of new binaries, among many possible outcomes. A few possibilities are flybys, exchanges, ionisation of a binary and triple formations. By denoting with $f_\text{\tiny X}$  the frequency a particular outcome X to occur, the corresponding process cross section between two hard binaries with mass components $m_1$, $m_2$ and $m_3$, $m_4$ respectively is given by \cite{Bacon:1996cj,Antognini:2015ima}
	\begin{equation}
	\Sigma_\text{\tiny X}=\pi f_\text{\tiny X}\max(a_{12},a_{34})^2\lp 4{v_c\over v_{12,34}}+3\rp^2,
	\end{equation}
	where we have defined
	\begin{equation}
	v_c=\left[{GM_{1234}\over M_{12}M_{34}}\lp{m_1m_2\over a_{12}}+{m_3m_4\over a_{34}}\rp\right]^{1/2}.
	\end{equation}
	Here, $v_{12,34}$ is the relative velocity between the 1-2 and 3-4 pairs at infinity, and $a_{12}$ and $a_{34}$ are their SMA in the pre-encounter state respectively.
	
	The possible frequency of a given outcome to occur will depend on the characteristic masses, velocities and semi-major axes of the binary under consideration. In order to have an estimate of the characteristic value of $f_\text{\tiny X}$ for the binary exchange, we have performed $10^4$ binary-binary scattering experiments using the numerical toolkit FEWBODY\footnote{The numerical toolkit FEWBODY is publicly available at \href{https://sourceforge.net/projects/fewbody/files/}{https://sourceforge.net/projects/fewbody/files/}.} \cite{Fregeau:2004if} with an initial state (80-10)-(60-10) and find that the frequency at which events result in the final state (80-60)-10-10 is of the order of $f_\text{\tiny X}=(32\pm 5)\%$. We will therefore adopt the value of $f_\text{\tiny X} \simeq 0.3$ in our computation as a characteristic value for the binary exchange to occur.
	
	The velocity averaged differential exchange rate per unit volume is then given by
	\begin{equation}
	\gamma_\text{\tiny X}=n_{12}n_{34}f_\text{\tiny X}\pi \max(a_{12},a_{34})^2\lp 16\sqrt{6\over\pi}{v_c^2\over\sigma_{12,34}}+24v_c+18\sqrt{2\over3\pi}\sigma_{12,34}\rp,
	\end{equation}
	where $\sigma_{12,34}$ is the three dimensional rms relative velocity between the 1-2 and 3-4 pairs at infinity. 
	We stress that a factor of 1/2 has been kept into account in the number densities when the same type of populations are considered in the interaction.
	The total exchange rate is finally given by multiplying the differential exchange rate per unit volume with the common segregation volume of the two binaries
	\begin{equation}
	\Gamma_\text{\tiny X}={4\pi\over3}\min\llp R_\text{\tiny seg}^3(M_{12}),R_\text{\tiny seg}^3(M_{34})\rrp \gamma_\text{\tiny X} .
	\end{equation}
	The SMA of the newly formed pair through a binary-binary interaction will depend on the intrinsic properties of the binaries under consideration. Instead of relying on a dedicated Monte Carlo simulation of the binary-binary interaction dynamics, 
	in our computation we have used Eq.~\ref{postSMA} to estimate the SMA of the new binary by considering the heavier binary (with a higher mass ratio) to play the role of the single object and interact with the other pair similarly to a binary-single interaction.

	\section{Appendix: The Catalysis Algorithm}
	\label{algorithm_appendix}
	\renewcommand{\theequation}{B.\arabic{equation}}
		\noindent
We develop a semi-analytical calculation of the corresponding merger rates induced from the ``catalysis" model. This is achieved by following a hierarchical process through which the population number of various binary species is calculated during a specific step and do not mix with those that have been produced from previous steps. This assumption allows to simplify the problem at hand and provide a conservative lower limit for the populations merger rates.
In the following, we present the algorithmic process used in solving the system in Fig.~\ref{sketch} in terms of a series of steps.
We stress that, in order to account for the distributions of masses and concentrations of the GCs, we perform an average over all the MWGCs and obtain the final merger rate density as
\be
R^\text{\tiny MWGC} = n_\text{\tiny GC} \langle \Gamma \rangle_\text{\tiny MWGC},
\ee
as defined in the text.

\begin{itemize}
\item One can start by identifying the initial sizes of the 10-10 binary population $N_\text{\tiny 10-10}^\text{\tiny (1)}$ and the single 60 (80) BH populations $N_\text{\tiny 60}^\text{\tiny (1)}$ ($N_\text{\tiny 80}^\text{\tiny (1)}$) in each cluster. The merger rate of 10-10 binaries can be computed from the fraction of those that can merge before interacting with a single 60 or 80 BHs, which is estimated according to Eq.~\eqref{eq: hardrate} as 
\be
\Gamma_\text{\tiny 10-10}^\text{\tiny (1)}=N_\text{\tiny 10-10}^\text{\tiny (1)} \frac{f_e(T_\text{\tiny 10-10})}{T_\text{\tiny 10-10}} \Theta(T_\text{\tiny cl} - T_\text{\tiny 10-10}),
\ee
in the characteristic timescale  $T_\text{\tiny 10-10}=\min\lp T_\text{\tiny cl},\Gamma_\text{\tiny A}^{-1}N_\text{\tiny 60}^\text{\tiny (1)},\Gamma_\text{\tiny B}^{-1}N_\text{\tiny 80}^\text{\tiny (1)}\rp$. The corresponding number of 60-10 and 80-10 pairs that can form through binary-single interactions (processes A and B in Fig.~\ref{sketch}) is given by
\begin{align}
N_\text{\tiny 60-10}=\min\left({\Gamma_\text{\tiny A}\over\Gamma_\text{\tiny A}+\Gamma_\text{\tiny B}+\Gamma_\text{\tiny 10-10}^\text{\tiny (1)}}N_\text{\tiny 10-10}^\text{\tiny (1)},N_\text{\tiny 60}^\text{\tiny (1)}\right), \nonumber \\
N_\text{\tiny 80-10}=\min\left({\Gamma_\text{\tiny B}\over\Gamma_\text{\tiny A}+\Gamma_\text{\tiny B}+\Gamma_\text{\tiny 10-10}^\text{\tiny (1)}}N_\text{\tiny 10-10}^\text{\tiny (1)},N_\text{\tiny 80}^\text{\tiny (1)}\right).
\end{align}
The total merger rate of 10-10 pairs is then obtained by adding the contribution from those 10-10 binaries that have not merged before $T_\text{\tiny 10-10}$ to those that were not exchanged by a single 60 or 80 as
\begin{align}
\Gamma_\text{\tiny 10-10}=\Gamma_\text{\tiny 10-10}^\text{\tiny (1)}+\lp N_\text{\tiny 10-10}^\text{\tiny (1)}-N_\text{\tiny 60-10}-N_\text{\tiny 80-10}-\Gamma_\text{\tiny 10-10}^\text{\tiny (1)}T_\text{\tiny 10-10}\rp {f_e(T_\text{\tiny cl})\over T_\text{\tiny cl}},
\end{align}
where the last term in the bracket accounts for the expected number of 10-10 binaries that have merged within $T_\text{\tiny 10-10}$.
This total rate, averaged over the MWGC catalog, corresponds to the red line in Fig.~\ref{Final_results}.

\item 
Single 60 and 80 objects in the cluster, with population number $N_\text{\tiny 60}^\text{\tiny (2)}=N_\text{\tiny 60}^\text{\tiny (1)}-N_\text{\tiny 60-10}$ and $N_\text{\tiny 80}^\text{\tiny (2)}=N_\text{\tiny 80}^\text{\tiny (1)}-N_\text{\tiny 80-10}$, can participate in binary-single interactions with the 60-10 and 80-10 pairs, which were formed within timescales given by $T^\text{\tiny form}_\text{\tiny 60-10}=\Gamma_\text{\tiny A}^{-1}N_\text{\tiny 60-10}$ and $T^\text{\tiny form}_\text{\tiny 80-10}=\Gamma_\text{\tiny B}^{-1}N_\text{\tiny 80-10}$, respectively. Those interactions are graphically represented in Fig.~\ref{sketch} by the C, D, E and G arrows, taking into account both the contribution of binary-single and binary-binary interactions (see Appendix.~\ref{exchanges_appendix} for details on the exchange rates expressions), leading to the formation of 60-60, 80-80 and 80-60 binaries.
The merger rates of the remaining 60-10 and 80-10 binaries are then given by 
\begin{align}
\Gamma_\text{\tiny 60-10}
=N_\text{\tiny 60-10}\frac{f_e(T_\text{\tiny 60-10})}{T_\text{\tiny 60-10}} \Theta(T_\text{\tiny cl} -T_\text{\tiny 60-10}), \qquad
\Gamma_\text{\tiny 80-10}
=N_\text{\tiny 80-10} \frac{f_e(T_\text{\tiny 80-10})}{T_\text{\tiny 80-10}} \Theta(T_\text{\tiny cl} -T_\text{\tiny 80-10}),
\end{align}
in terms of the characteristic timescales $T_\text{\tiny 60-10}=\min\lp T_\text{\tiny cl}-T^\text{\tiny form}_\text{\tiny 60-10},\Gamma_\text{\tiny C}^{-1}N_\text{\tiny 60-10},\Gamma_\text{\tiny D}^{-1}N_\text{\tiny 60-10}\rp$ and
$T_\text{\tiny 80-10}=\min\lp T_\text{\tiny cl}-T^\text{\tiny form}_\text{\tiny 80-10},\Gamma_\text{\tiny E}^{-1}N_\text{\tiny 80-10},\Gamma_\text{\tiny G}^{-1}N_\text{\tiny 80-10}\rp$, where the delay in the formation of the binary populations  has been properly taken into account. These rates, averaged over the MWGC catalog, correspond to the yellow and green curves  in Fig.~\ref{Final_results}.

\item From the predicted population number of 80-80 and 60-60 binaries, along with their characteristic formation timescales, given by 
\begin{align}
N_\text{\tiny 80-80}&={\Gamma_\text{\tiny G}\over\Gamma_\text{\tiny E}+\Gamma_\text{\tiny G}+\Gamma_\text{\tiny 80-10}}N_\text{\tiny 80-10}, \qquad 
T^\text{\tiny form}_\text{\tiny 80-80}=T^\text{\tiny form}_\text{\tiny 80-10}+\Gamma_\text{\tiny G}^{-1}N_\text{\tiny 80-80}, \nonumber \\
N_\text{\tiny 60-60}&={\Gamma_\text{\tiny C}\over\Gamma_\text{\tiny D}+\Gamma_\text{\tiny C}+\Gamma_\text{\tiny 60-10}}N_\text{\tiny 60-10}, \qquad 
T^\text{\tiny form}_\text{\tiny 60-60}=T^\text{\tiny form}_\text{\tiny 60-10}+\Gamma_\text{\tiny C}^{-1}N_\text{\tiny 60-60},
\end{align}
one can compute their expected merger rates as
\begin{align}
\Gamma_\text{\tiny 80-80}
=N_\text{\tiny 80-80} \frac{f_e(T_\text{\tiny 80-80})}{T_\text{\tiny 80-80}} \Theta(T_\text{\tiny cl} -T_\text{\tiny 80-80}), \qquad 
\Gamma_\text{\tiny 60-60}
=N_\text{\tiny 60-60} \frac{f_e(T_\text{\tiny 60-60})}{T_\text{\tiny 60-60}} \Theta(T_\text{\tiny cl} -T_\text{\tiny 60-60}),
\end{align}
in terms of the characteristic timescales  $T_\text{\tiny 80-80}=T_\text{\tiny cl}-T^\text{\tiny form}_\text{\tiny 80-80}$ and $T_\text{\tiny 60-60}=T_\text{\tiny cl}-T^\text{\tiny form}_\text{\tiny 60-60}$. 
These rates, averaged over the MWGC catalog, correspond to the magenta and blue  curves  in Fig.~\ref{Final_results}.

We treat the 80-60 population differently by separating its merger rate into the three components which correspond to the three different ways of synthesizing 80-60 pairs through D$_\text{s}$, E$_\text{s}$ and D$_\text{\tiny b}$ (processes D$_\text{\tiny b}$ and E$_\text{\tiny b}$ are identical), i.e. $\Gamma_\text{\tiny 80-60}=\Gamma_\text{\tiny 80-60}^{\text{\tiny D}_\text{s}}+\Gamma_\text{\tiny 80-60}^{\text{\tiny E}_\text{s}}+\Gamma_\text{\tiny 80-60}^{\text{\tiny D}_\text{\tiny b}}$. The former two channels, which originate from binary-single interactions, are characterised by the population numbers and formation timescales 
\begin{align}
N_\text{\tiny 80-60}^{\text{\tiny D}_\text{s}}&={\Gamma_{\text{\tiny D}_\text{s}}\over\Gamma_\text{\tiny C}+\Gamma_\text{\tiny D}+\Gamma_\text{\tiny 60-10}}N_\text{\tiny 60-10}, \qquad T^{\text{\tiny form}(\text{\tiny D}_\text{s})}_\text{\tiny 80-60}=T^\text{\tiny form}_\text{\tiny 60-10}+{N_\text{\tiny 80-60}^{\text{\tiny D}_\text{s}}\over\Gamma_\text{\tiny 80-60}^{\text{\tiny D}_\text{s}}},
 \\
 N_\text{\tiny 80-60}^{\text{\tiny E}_\text{s}}&={\Gamma_{\text{\tiny E}_\text{s}}\over\Gamma_\text{\tiny G}+\Gamma_\text{\tiny E}+\Gamma_\text{\tiny 80-10}}N_\text{\tiny 80-10}, \qquad 
T^{\text{\tiny form}(\text{\tiny E}_\text{s})}_\text{\tiny 80-60}=T^\text{\tiny form}_\text{\tiny 80-10}+{N_\text{\tiny 80-60}^{\text{\tiny E}_\text{s}}\over\Gamma_\text{\tiny 80-60}^{\text{\tiny E}_\text{s}}},
\end{align}
such that the merger rates are easily computed using the known hardening formula in Eq.~\eqref{eq: hardrate},
in terms of the characteristic merger timescales $T_\text{\tiny 80-60}^{\text{\tiny D}_\text{s}}=T_\text{\tiny cl}-T^{\text{\tiny form}(\text{\tiny D}_\text{s})}_\text{\tiny 80-60}$ and $T_\text{\tiny 80-60}^{\text{\tiny E}_\text{s}}=T_\text{\tiny cl}-T^{\text{\tiny form}(\text{\tiny E}_\text{s})}_\text{\tiny 80-60}$.
The latter channel, which originates from binary-binary interaction, is instead characterised by the population number 
\begin{align}
N_\text{\tiny 80-60}^{\text{\tiny D}_\text{\tiny b}}=\min\left[\max\lp{\Gamma_{\text{\tiny D}_\text{\tiny b}}N_\text{\tiny 60-10}\over\Gamma_\text{\tiny D}+\Gamma_\text{\tiny C}+\Gamma_\text{\tiny 60-10}},{\Gamma_{\text{\tiny E}_\text{\tiny b}}N_\text{\tiny 80-10}\over\Gamma_\text{\tiny E}+\Gamma_\text{\tiny G}+\Gamma_\text{\tiny 80-10}}\rp,N_\text{\tiny 60-10},N_\text{\tiny 80-10}\right],
\end{align}
which properly accounts for the largest fraction of 60-10 or 80-10 pairs that contributes into 80-60 via binary-binary interactions, with the additional request that the number of 80-60 pairs does not exceed either the initial number of 60-10 or 80-10 binaries.
The formation timescale is given by $T^{\text{\tiny form}(\text{\tiny D}_\text{\tiny b})}_\text{\tiny 80-60}=\max\lp T^\text{\tiny form}_\text{\tiny 60-10},T^\text{\tiny form}_\text{\tiny 80-10}\rp+\lp\Gamma_\text{\tiny 80-60}^{\text{\tiny D}_\text{\tiny b}}\rp^{-1}N_\text{\tiny 80-60}^{\text{\tiny D}_\text{\tiny b}}$ and the merger timescale by $T_\text{\tiny 80-60}^{\text{\tiny D}_\text{b}}=T_\text{\tiny cl}-T^{\text{\tiny form}(\text{\tiny D}_\text{b})}_\text{\tiny 80-60}$.
Finally, the total merger rate of the 80-60 binaries is  given by
\begin{align}
\Gamma_\text{\tiny 80-60}&=
N_\text{\tiny 80-60}^{\text{\tiny D}_\text{s}}{f_e\lp T_\text{\tiny 80-60}^{\text{\tiny D}_\text{s}}\rp\over T_\text{\tiny 80-60}^{\text{\tiny D}_\text{s}}} \Theta(T_\text{\tiny cl} -T^{\text{\tiny D}_\text{s}}_\text{\tiny 80-60}) +
N_\text{\tiny 80-60}^{\text{\tiny E}_\text{s}}{f_e\lp T_\text{\tiny 80-60}^{\text{\tiny E}_\text{s}}\rp\over T_\text{\tiny 80-60}^{\text{\tiny E}_\text{s}}} \Theta(T_\text{\tiny cl} -T^{\text{\tiny E}_\text{s}}_\text{\tiny 80-60}) 
\nonumber \\
& +  N_\text{\tiny 80-60}^{\text{\tiny D}_\text{b}}{f_e\lp T_\text{\tiny 80-60}^{\text{\tiny D}_\text{b}}\rp\over T_\text{\tiny 80-60}^{\text{\tiny D}_\text{b}}} \Theta(T_\text{\tiny cl} -T^{\text{\tiny D}_\text{b}}_\text{\tiny 80-60}),
\end{align}
which is plotted, once averaged over the MWGC catalog, as the black dashed line in Fig.~\ref{Final_results}.

\end{itemize}

\section{Appendix: LIGO/Virgo sensitivity and observable rates}
\label{sensitivity_appendix}
\renewcommand{\theequation}{C.\arabic{equation}}
\noindent
A BH binary is typically characterized by the component masses $m_1$ and $m_2$, 
dimensionless spins $\boldsymbol{\chi}_1$ and $\boldsymbol{\chi}_2$ and source redshift $z$. 
When one considers the measurement process, each individual binary is accompanied by extrinsic parameters related to the position and orientation with respect to the detectors. Those are defined in terms of right ascension $\alpha$, declination $\delta$, orbital-plane inclination $\iota$, and polarization angle $\psi$.  
One can therefore define the intrinsic and extrinsic binary parameters as
$\theta=\{m1,m2,\boldsymbol{\chi_1},\boldsymbol{\chi_2}\}$ 
and $\lambda = \{\alpha,\delta,\iota,\psi\}$
respectively. We will neglect the role of the spin in the following estimates.

In order to compute the probability of detection of a binary at the experiments, one has to marginalize the detectability over the extrinsic parameters $\lambda$.  For each value of the intrinsic parameter set $\theta$, one can define the detection probability as
\begin{equation}
p_{\rm det}(\theta)= \int  p(\lambda) \, \Theta[\rho(\theta, \lambda) - \rho_{\rm thr}] \, d\lambda\,,
\label{pdetdefine}
\end{equation}
where $p(\lambda)$ is the probability distribution function of $\lambda$ and $\rho$ is the Signal-to-Noise Ratio (SNR).  For a single detector, the threshold for detection is typically taken  to be $\rho_{\rm thr}=8$.
One can extract the dependency of the SNR on $\lambda$ and factorise  $\rho(\theta,\lambda)=\omega(\lambda) \rho_\text{\tiny opt}(\theta)$, where $\rho_\text{\tiny opt}$ is the SNR of an ``optimal'' source 
\begin{equation}
\rho_\text{\tiny opt}^2 (m_1,m_2,z)\equiv 4 \int_0^\infty \frac{ |\tilde h (\nu)|^2}{S_n (\nu)} {\rm d} \nu,
\end{equation}
in terms of the waveform $\tilde h(\nu)$ in Fourier space and the  strain noise $S_n (\nu)$ of the detector.
In the following we will estimate the SNR using the
strain as measured during the O3 run of the LIGO/Virgo experiment.
Finally, one can compute the  marginalized distribution $p_{\rm det}(\theta)$ by calculating the integral $P(\omega)=\int_\omega^1 p(\omega') d\omega'$ at  $\omega=\rho_{\rm thr}/\rho_\text{\tiny opt}(\theta)$. %
In the case of isotropic sources, $\alpha$, $\cos\delta$, $\cos\iota$, and $\psi$ are uniformly distributed and the result of the integral 
gives rise to the function $P(\omega)$ as found in Ref.~\cite{berti}.

The detection probability can be linked to the ``effective'' spacetime volume $\langle V T_{\rm obs} \rangle$ by %
\begin{equation}
\langle V T_{\rm obs} \rangle_{m_1\text{-}m_2} = T_{\rm obs} \int p_{\rm det}(\theta) \frac{dV_c}{dz} \frac{1}{1+z}  dz \,,
\end{equation}
where $T_{\rm obs}$ is the duration of the observing run and $V_c$ is the comoving volume.

Assuming the merger rate $R$ to be constant within the observable redshift horizon, the expected number of detections for a binary ${m_1\text{-}m_2}$ can be found as 
$
N_{m_1\text{-}m_2} = \langle V T_\text{\tiny obs} \rangle_{m_1\text{-}m_2} R.
$
In particular, during the LIGO/Virgo third observing run, one finds
\begin{align}
\langle V T_\text{\tiny obs} \rangle _\text{80-60} &= 12\ \text{Gpc}^{3} \text{yr},
\\
\langle V T_\text{\tiny obs} \rangle _\text{80-10} &= 0.69\ \text{Gpc}^{3} \text{yr}. 
\end{align}
As one can appreciate, the detector selection bias makes harder to observe un-equal mass binaries.
Therefore, even though the merger rates of 80-10 and 80-60 are comparable in most of the parameter space of the ``catalysis'' (see Fig.~\ref{Final_results}), in this scenario one does not expect asymmetric ABH-PBH binaries in the LIGO/Virgo current GWTC-2 catalog events.



\begin{thebibliography}{99}



\bibitem{TheLIGOScientific:2014jea}
J.~Aasi \textit{et al.} [LIGO Scientific],
Class. Quant. Grav. \textbf{32} (2015), 074001
\arXiv{1411.4547}{gr-qc}.

\bibitem{Bird:2016dcv}
S.~Bird, I.~Cholis, J.~B.~Mu\~noz, Y.~Ali-Ha\"\i{}moud, M.~Kamionkowski, E.~D.~Kovetz, A.~Raccanelli and A.~G.~Riess,
Phys. Rev. Lett. \textbf{116} (2016) no.20, 201301
\arXiv{1603.00464}{astro-ph.CO}.

\bibitem{Sasaki:2016jop}
M.~Sasaki, T.~Suyama, T.~Tanaka and S.~Yokoyama,
Phys. Rev. Lett. \textbf{117} (2016) no.6, 061101
[erratum: Phys. Rev. Lett. \textbf{121} (2018) no.5, 059901]
\arXiv{1603.08338}{astro-ph.CO}.

\bibitem{Blinnikov:2016bxu}
S.~Blinnikov, A.~Dolgov, N.~K.~Porayko and K.~Postnov,
JCAP \textbf{11} (2016), 036
\arXiv{1611.00541}{astro-ph.HE}.

\bibitem{Clesse:2016vqa}
S.~Clesse and J.~Garc\'\i{}a-Bellido,
Phys. Dark Univ. \textbf{15} (2017), 142-147
\arXiv{1603.05234}{astro-ph.CO}.

\bibitem{Nakamura:1997sm}
T.~Nakamura, M.~Sasaki, T.~Tanaka and K.~S.~Thorne,
Astrophys. J. Lett. \textbf{487} (1997), L139-L142
\arXivold{astro-ph/9708060}.

\bibitem{Raidal:2017mfl}
M.~Raidal, V.~Vaskonen and H.~Veerm\"ae,
JCAP \textbf{09} (2017), 037
\arXiv{1707.01480}{astro-ph.CO}.

 \bibitem{Ali-Haimoud:2017rtz}
 Y.~Ali-Ha\"\i{}moud, E.~D.~Kovetz and M.~Kamionkowski,
 Phys. Rev. D \textbf{96} (2017) no.12, 123523
 \arXiv{1709.06576}{astro-ph.CO}.


 \bibitem{raid}
 M.~Raidal, C.~Spethmann, V.~Vaskonen and H.~Veerm\"ae,
 JCAP \textbf{02} (2019), 018
 \arXiv{1812.01930}{astro-ph.CO}.
 
 
\bibitem{Inman:2019wvr}
D.~Inman and Y.~Ali-Haïmoud,
Phys. Rev. D \textbf{100} (2019) no.8, 083528
\arXiv{1907.08129}{astro-ph.CO}.


\bibitem{Liu:2019rnx}
L.~Liu, Z.~K.~Guo and R.~G.~Cai,
Eur. Phys. J. C \textbf{79} (2019) no.8, 717
\arXiv{1901.07672}{astro-ph.CO}.

\bibitem{Vaskonen:2019jpv}
V.~Vaskonen and H.~Veerm\"ae,
Phys. Rev. D \textbf{101} (2020) no.4, 043015
\arXiv{1908.09752}{astro-ph.CO}.


\bibitem{Gow:2019pok}
A.~D.~Gow, C.~T.~Byrnes, A.~Hall and J.~A.~Peacock,
JCAP \textbf{01} (2020), 031
\arXiv{1911.12685}{astro-ph.CO}.

\bibitem{Dolgov:2020xzo}
A.~D.~Dolgov, A.~G.~Kuranov, N.~A.~Mitichkin, S.~Porey, K.~A.~Postnov, O.~S.~Sazhina and I.~V.~Simkin,
\arXiv{2005.00892}{astro-ph.CO}.

\bibitem{DeLuca:2020qqa}
V.~De Luca, G.~Franciolini, P.~Pani and A.~Riotto,
JCAP \textbf{06} (2020), 044
\arXiv{2005.05641}{astro-ph.CO}.

\bibitem{Jedamzik:2020ypm}
K.~Jedamzik,
JCAP \textbf{09} (2020), 022
\arXiv{2006.11172}{astro-ph.CO}.

\bibitem{Jedamzik:2020omx}
K.~Jedamzik,
Phys. Rev. Lett. \textbf{126} (2021) no.5, 051302
\arXiv{2007.03565}{astro-ph.CO}.

\bibitem{Clesse:2020ghq}
S.~Clesse and J.~Garcia-Bellido,
\arXiv{2007.06481}{astro-ph.CO}.


\bibitem{Hall:2020daa}
A.~Hall, A.~D.~Gow and C.~T.~Byrnes,
Phys. Rev. D \textbf{102} (2020), 123524
\arXiv{2008.13704}{astro-ph.CO}.



\bibitem{DeLuca:2020jug}
V.~De Luca, V.~Desjacques, G.~Franciolini and A.~Riotto,
JCAP \textbf{11} (2020), 028
\arXiv{2009.04731}{astro-ph.CO}.

\bibitem{Tkachev:2020uin}
M.~Tkachev, S.~Pilipenko and G.~Yepes,
Mon. Not. Roy. Astron. Soc. \textbf{499} (2020) no.4, 4854-4862
\arXiv{2009.07813}{astro-ph.CO}.

\bibitem{Wong:2020yig}
K.~W.~K.~Wong, G.~Franciolini, V.~De Luca, V.~Baibhav, E.~Berti, P.~Pani and A.~Riotto,
Phys. Rev. D \textbf{103} (2021) no.2, 023026
\arXiv{2011.01865}{gr-qc}.

\bibitem{Hutsi:2020sol}
G.~H\"utsi, M.~Raidal, V.~Vaskonen and H.~Veerm\"ae,
JCAP \textbf{03} (2021), 068
\arXiv{2012.02786}{astro-ph.CO}.


		\bibitem{Abbott:2020gyp}
		R.~Abbott \textit{et al.} [LIGO Scientific and Virgo],
		\arXiv{2010.14533}{astro-ph.HE}.
				
		\bibitem{Abbott:2020niy}
		R.~Abbott \textit{et al.} [LIGO Scientific and Virgo],
\arXiv{2010.14527}{gr-qc}


\bibitem{Abbott:2020mjq}
R.~Abbott \textit{et al.} [LIGO Scientific and Virgo],
Astrophys. J. \textbf{900} (2020) no.1, L13
\arXiv{2009.01190}{astro-ph.HE}.

\bibitem{Abbott:2020tfl}
R.~Abbott \textit{et al.} [LIGO Scientific and Virgo],
Phys. Rev. Lett. \textbf{125} (2020) no.10, 101102
\arXiv{2009.01075}{gr-qc}.


\bibitem{DeLuca:2020sae}
V.~De Luca, V.~Desjacques, G.~Franciolini, P.~Pani and A.~Riotto,
Phys. Rev. Lett. \textbf{126} (2021) no.5, 051101
\arXiv{2009.01728}{astro-ph.CO}.



\bibitem{Antonini:2018auk}
F.~Antonini, M.~Gieles and A.~Gualandris,
Mon. Not. Roy. Astron. Soc. \textbf{486} (2019) no.4, 5008-5021
\arXiv{1811.03640}{astro-ph.HE}.

\bibitem{Rodriguez:2019huv}
C.~L.~Rodriguez, M.~Zevin, P.~Amaro-Seoane, S.~Chatterjee, K.~Kremer, F.~A.~Rasio and C.~S.~Ye,
Phys. Rev. D \textbf{100} (2019) no.4, 043027
\arXiv{1906.10260}{astro-ph.HE}.

\bibitem{Gerosa:2019zmo}
D.~Gerosa and E.~Berti,
Phys. Rev. D \textbf{100} (2019) no.4, 041301
\arXiv{1906.05295}{astro-ph.HE}.

\bibitem{Baibhav:2020xdf}
V.~Baibhav, D.~Gerosa, E.~Berti, K.~W.~K.~Wong, T.~Helfer and M.~Mould,
Phys. Rev. D \textbf{102} (2020) no.4, 043002
\arXiv{2004.00650}{astro-ph.HE}.

\bibitem{Mapelli:2020xeq}
M.~Mapelli, F.~Santoliquido, Y.~Bouffanais, M.~A.~Sedda, N.~Giacobbo, M.~C.~Artale and A.~Ballone,
\arXiv{2007.15022}{astro-ph.HE}.

\bibitem{Liu:2020gif}
B.~Liu and D.~Lai,
Mon. Not. Roy. Astron. Soc. \textbf{502} (2021) no.2, 2049-2064
\arXiv{2009.10068}{astro-ph.HE}.

\bibitem{Fragione:2020han}
G.~Fragione, A.~Loeb and F.~A.~Rasio,
Astrophys. J. Lett. \textbf{902} (2020) no.1, L26
\arXiv{2009.05065}{astro-ph.GA}.

\bibitem{Anagnostou:2020umw}
O.~Anagnostou, M.~Trenti and A.~Melatos,
\arXiv{2010.06161}{astro-ph.HE}.

\bibitem{Kimball:2020qyd}
C.~Kimball, C.~Talbot, C.~P.~L.~Berry, M.~Zevin, E.~Thrane, V.~Kalogera, R.~Buscicchio, M.~Carney, T.~Dent and H.~Middleton, \textit{et al.}
\arXiv{2011.05332}{astro-ph.HE}.

\bibitem{Kremer:2020wtp}
K.~Kremer, M.~Spera, D.~Becker, S.~Chatterjee, U.~N.~Di Carlo, G.~Fragione, C.~L.~Rodriguez, C.~S.~Ye and F.~A.~Rasio,
Astrophys. J. \textbf{903} (2020) no.1, 45
\arXiv{2006.10771}{astro-ph.HE}.


\bibitem{Liu:2020lmi}
B.~Liu and V.~Bromm,
Astrophys. J. Lett. \textbf{903} (2020) no.2, L40
\arXiv{2009.11447}{astro-ph.GA}.

\bibitem{Kinugawa:2020xws}
T.~Kinugawa, T.~Nakamura and H.~Nakano,
Mon. Not. Roy. Astron. Soc. \textbf{501} (2021) no.1, L49-L53
\arXiv{2009.06922}{astro-ph.HE}.

\bibitem{Safarzadeh:2020vbv}
M.~Safarzadeh and Z.~Haiman,
Astrophys. J. Lett. \textbf{903} (2020) no.1, L21
\arXiv{2009.09320}{astro-ph.HE}.

\bibitem{Belczynski:2020bca}
K.~Belczynski,
Astrophys. J. Lett. \textbf{905} (2020) no.2, L15
\arXiv{2009.13526}{astro-ph.HE}.

\bibitem{Gayathri:2020coq}
V.~Gayathri, J.~Healy, J.~Lange, B.~O'Brien, M.~Szczepanczyk, I.~Bartos, M.~Campanelli, S.~Klimenko, C.~Lousto and R.~O'Shaughnessy,
\arXiv{2009.05461}{astro-ph.HE}.

\bibitem{Romero-Shaw:2020thy}
I.~M.~Romero-Shaw, P.~D.~Lasky, E.~Thrane and J.~C.~Bustillo,
Astrophys. J. Lett. \textbf{903} (2020) no.1, L5
\arXiv{2009.04771}{astro-ph.HE}.


\bibitem{Gondan:2020svr}
L.~Gond\'an and B.~Kocsis,
\arXiv{2011.02507}{astro-ph.HE}.


\bibitem{Fishbach:2020qag}
M.~Fishbach and D.~E.~Holz,
Astrophys. J. Lett. \textbf{904} (2020) no.2, L26
\arXiv{2009.05472}{astro-ph.HE}.

\bibitem{Nitz:2020mga}
A.~H.~Nitz and C.~D.~Capano,
Astrophys. J. Lett. \textbf{907} (2021) no.1, L9
\arXiv{2010.12558}{astro-ph.HE}.

\bibitem{Samsing:2020tda}
J.~Samsing, I.~Bartos, D.~J.~D'Orazio, Z.~Haiman, B.~Kocsis, N.~W.~C.~Leigh, B.~Liu, M.~E.~Pessah and H.~Tagawa,
\arXiv{2010.09765}{astro-ph.HE}.

\bibitem{Tagawa:2020jnc}
H.~Tagawa, B.~Kocsis, Z.~Haiman, I.~Bartos, K.~Omukai and J.~Samsing,
Astrophys. J. Lett. \textbf{907} (2021) no.1, L20
\arXiv{2010.10526}{astro-ph.HE}.

\bibitem{Sasaki:2018dmp}
M.~Sasaki, T.~Suyama, T.~Tanaka and S.~Yokoyama,
Class. Quant. Grav. \textbf{35} (2018) no.6, 063001
\arXiv{1801.05235}{astro-ph.CO}.

\bibitem{Carr:2020gox}
B.~Carr, K.~Kohri, Y.~Sendouda and J.~Yokoyama,
\arXiv{2002.12778}{astro-ph.CO}.

\bibitem{green} 
A.~M.~Green and B.~J.~Kavanagh,
J. Phys. G \textbf{48} (2021) no.4, 4
\arXiv{2007.10722}{astro-ph.CO}.

\bibitem{Rodriguez:2015oxa}
C.~L.~Rodriguez, M.~Morscher, B.~Pattabiraman, S.~Chatterjee, C.~J.~Haster and F.~A.~Rasio,
Phys. Rev. Lett. \textbf{115} (2015) no.5, 051101
[erratum: Phys. Rev. Lett. \textbf{116} (2016) no.2, 029901]
\arXiv{1505.00792}{astro-ph.HE}.

\bibitem{Rodriguez:2016kxx}
C.~L.~Rodriguez, S.~Chatterjee and F.~A.~Rasio,
Phys. Rev. D \textbf{93} (2016) no.8, 084029
\arXiv{1602.02444}{astro-ph.HE}.

\bibitem{OLeary:2005vqo}
R.~M.~O'Leary, F.~A.~Rasio, J.~M.~Fregeau, N.~Ivanova and R.~W.~O'Shaughnessy,
Astrophys. J. \textbf{637} (2006), 937-951
\arXivold{astro-ph/0508224}.

\bibitem{Miller:2008yw}
M.~C.~Miller and V.~M.~Lauburg,
Astrophys. J. \textbf{692} (2009), 917-923
\arXiv{0804.2783}{astro-ph}.

\bibitem{Antonini:2016gqe}
F.~Antonini and F.~A.~Rasio,
Astrophys. J. \textbf{831} (2016) no.2, 187
\arXiv{1606.04889}{astro-ph.HE}.

\bibitem{Nitz:2020bdb}
A.~H.~Nitz and Y.~F.~Wang,
Phys. Rev. Lett. \textbf{126} (2021) no.2, 021103
\arXiv{2007.03583}{astro-ph.HE}.

\bibitem{Tsai:2020hpi}
Y.~D.~Tsai, A.~Palmese, S.~Profumo and T.~Jeltema,
\arXiv{2007.03686}{astro-ph.HE}.

\bibitem{Vattis:2020iuz}
K.~Vattis, I.~S.~Goldstein and S.~M.~Koushiappas,
Phys. Rev. D \textbf{102} (2020) no.6, 061301
\arXiv{2006.15675}{astro-ph.HE}.

\bibitem{Kovetz:2018vly}
E.~D.~Kovetz, I.~Cholis, M.~Kamionkowski and J.~Silk,
Phys. Rev. D \textbf{97} (2018) no.12, 123003
\arXiv{1803.00568}{astro-ph.HE}.

\bibitem{Kritos:2020fjw}
K.~Kritos and I.~Cholis,
Phys. Rev. D {\bf 102} (2020), 083016
\arXiv{2007.02968}{astro-ph.GA}.


\bibitem{harris2010new}
Harris, W.E. 1996,
AJ, \textbf{112}, 1487,
\arXiv{1012.3224}{astro-ph.GA}

\bibitem{King1962}
I.~King, Astron. J. \textbf{67} (1962), 471

\bibitem{Chatterjee:2016hxc}
S.~Chatterjee, C.~L.~Rodriguez and F.~A.~Rasio,
Astrophys. J. \textbf{834} (2017) no.1, 68
\arXiv{1603.00884}{astro-ph.GA}.


\bibitem{Valcin:2020vav}
D.~Valcin, J.~L.~Bernal, R.~Jimenez, L.~Verde and B.~D.~Wandelt,
\arXiv{2007.06594}{astro-ph.CO}.

\bibitem{Flitter:2020bky}
J.~Flitter, J.~B.~Mu\~noz and E.~D.~Kovetz,
\arXiv{2008.10389}{astro-ph.HE}.

\bibitem{PortegiesZwart:1999nm}
S.~F.~Portegies Zwart and S.~McMillan,
Astrophys. J. Lett. \textbf{528} (2000), L17
\arXiv{astro-ph/9910061}{astro-ph}.

\bibitem{Antonini:2020xnd}
F.~Antonini and M.~Gieles,
Phys. Rev. D \textbf{102} (2020), 123016
\arXiv{2009.01861}{astro-ph.HE}.

\bibitem{1971ApJ...164..399S}
L.~Jr.~Spitzer and M.~Hart,
Astrophys. J. \textbf{164} (1971), 399.


\bibitem{Fragione:2020fnr}
G.~Fragione, R.~Perna and A.~Loeb,
Mon. Not. Roy. Astron. Soc. \textbf{500} (2020) no.4, 4307-4318
\arXiv{2006.14632}{astro-ph.GA}.

\bibitem{1987gady.book.....B}
J.~Binney and S.~Tremaine, ``Galactic dynamics'', 1987.

\bibitem{Khalisi:2006ne}
E.~Khalisi, P.~Amaro-Seoane and R.~Spurzem,
Mon. Not. Roy. Astron. Soc. \textbf{374} (2007), 703-720
\arXivold{astro-ph/0602570}.

\bibitem{Trenti:2013ety}
M.~Trenti and R.~van der Marel,
Mon. Not. Roy. Astron. Soc. \textbf{435} (2013), 3272
\arXiv{1302.2152}{astro-ph.GA}.

\bibitem{Bahcall:1976aa} 
J.~N.~Bahcall and R.~A.~Wolf,
Astrophys.\ J.\  {\bf 209}, 214 (1976).

\bibitem{cl1} Y.~Ali-Ha\"{i}moud,
Phys.\ Rev.\ Lett.\  {\bf 121}, no. 8, 081304 (2018)
\arXiv{1805.05912}{astro-ph.CO}.

\bibitem{cl2} V.~Desjacques and A.~Riotto,
Phys.\ Rev.\ D {\bf 98}, no. 12, 123533 (2018)
\arXiv{1806.10414}{astro-ph.CO}.

\bibitem{cl3} G.~Ballesteros, P.~D.~Serpico and M.~Taoso,
JCAP {\bf 1810}, 043 (2018)
\arXiv{1807.02084}{astro-ph.CO}.

\bibitem{cl4} A.~Moradinezhad Dizgah, G.~Franciolini and A.~Riotto,
JCAP {\bf 1911}, no. 11, 001 (2019)
\arXiv{1906.08978}{astro-ph.CO}.

\bibitem{Kroupa:2002ky}
P.~Kroupa,
Science \textbf{295} (2002), 82-91
\arXivold{astro-ph/0201098}.

\bibitem{Fryer:1999ht}
C.~L.~Fryer and V.~Kalogera,
Astrophys. J. \textbf{554} (2001), 548-560
\arXivold{astro-ph/9911312}.

\bibitem{Kovetz:2016kpi}
E.~D.~Kovetz, I.~Cholis, P.~C.~Breysse and M.~Kamionkowski,
Phys. Rev. D \textbf{95} (2017) no.10, 103010
\arXiv{1611.01157}{astro-ph.CO}.

\bibitem{catalog}
B.~P.~Abbott \textit{et al.} [LIGO Scientific and Virgo],
Astrophys. J. Lett. \textbf{882}, no.2, L24 (2019)
\arXiv{1811.12940}{astro-ph.HE}.


\bibitem{Pavlik2018}
V.~Pavlík, T.~Jeřábková, P.~Kroupa and H.Baumgardt,
AAP {\bf 617} (2018), A69
\arXiv{1806.05192}{astro-ph.GA}.

\bibitem{ser}
P.~D.~Serpico, V.~Poulin, D.~Inman and K.~Kohri,
Phys. Rev. Res. \textbf{2} (2020) no.2, 023204
\arXiv{2002.10771}{astro-ph.CO}.

\bibitem{DeLuca:2020fpg}
V.~De Luca, G.~Franciolini, P.~Pani and A.~Riotto,
Phys. Rev. D \textbf{102} (2020) no.4, 043505
\arXiv{2003.12589}{astro-ph.CO}.

\bibitem{2015ApJ...800....9M}
M.~Morscher, B.~Pattabiraman, C.~Rodriguez, F.~A.~Rasio and S.~Umbreit,
ApJ {\bf 800} (2015), 9

\bibitem{Fregeau:2001rp}
J.~M.~Fregeau, K.~J.~Joshi, S.~F.~Portegies Zwart and F.~A.~Rasio,
Astrophys. J. \textbf{570} (2002), 171-183
\arXivold{astro-ph/0111057}.

\bibitem{peters1964}
Peters, P.~C.,
Phys. Rev. \textbf{136} (1964), 1224-1232.

\bibitem{Sesana:2015haa}
A.~Sesana and F.~M.~Khan,
Mon. Not. Roy. Astron. Soc. \textbf{454}, no.1, L66-L70 (2015)
\arXiv{1505.02062}{astro-ph.GA}.

\bibitem{1975MNRAS.173..729H}
D.~C.~Heggie,
MNRAS {\bf 173} (1975), 729-787

\bibitem{1980AJ.....85.1281H}
J.~G.~Hills and L.~W.~Fullerton,
AJ {\bf 85} (1980), 1281-1291.


\bibitem{Samsing:2017xmd}
J.~Samsing,
Phys. Rev. D \textbf{97} (2018) no.10, 103014
\arXiv{1711.07452}{astro-ph.HE}.

\bibitem{Samsing:2017plz}
J.~Samsing and T.~Ilan,
Mon. Not. Roy. Astron. Soc. \textbf{476} (2018) no.2, 1548-1560
\arXiv{1706.04672}{astro-ph.HE}.

\bibitem{Samsing:2019dtb}
J.~Samsing, D.~J.~D'Orazio, K.~Kremer, C.~L.~Rodriguez and A.~Askar,
Phys. Rev. D \textbf{101} (2020) no.12, 123010
\arXiv{1907.11231}{astro-ph.HE}.

		\bibitem{Samsing:2020qqd}
		J.~Samsing and K.~Hotokezaka,
		\arXiv{2006.09744}{astro-ph.HE}.


\bibitem{1983ApJ...268..319H}
P.~Hut and J.~N.~Bahcall,
ApJ {\bf 268} (1983), 319-341


\bibitem{Sesana:2006xw}
A.~Sesana, F.~Haardt and P.~Madau,
Astrophys. J. \textbf{651} (2006), 392-400
\arXivold{astro-ph/0604299}.

\bibitem{Quinlan:1996vp}
G.~D.~Quinlan,
New Astron. \textbf{1} (1996), 35-56
\arXivold{astro-ph/9601092}.




\bibitem{TheLIGOScientific:2017qsa}
B.~P.~Abbott \textit{et al.} [LIGO Scientific and Virgo],
Phys. Rev. Lett. \textbf{119} (2017) no.16, 161101
\arXiv{1710.05832}{gr-qc}.


\bibitem
{quin}
G.~D.~Quinlan and S.~L.~Shapiro, ApJ {\bf 343} (1989), 725.

\bibitem{mouri}
H.~Mouri and Y.~Taniguchi,
Astrophys. J. Lett. \textbf{566} (2002), L17-L20
\arXivold{astro-ph/0201102}.

\bibitem{Cholis:2016kqi}
I.~Cholis, E.~D.~Kovetz, Y.~Ali-Ha\"\i{}moud, S.~Bird, M.~Kamionkowski, J.~B.~Mu\~noz and A.~Raccanelli,
Phys. Rev. D \textbf{94} (2016) no.8, 084013
\arXiv{1606.07437}{astro-ph.HE}.

\bibitem{OLeary:2008myb}
R.~M.~O'Leary, B.~Kocsis and A.~Loeb,
Mon. Not. Roy. Astron. Soc. \textbf{395} (2009) no.4, 2127-2146
\arXiv{0807.2638}{astro-ph}.


\bibitem{Ivanova:2005mi}
N.~Ivanova, K.~Belczynski, J.~M.~Fregeau and F.~A.~Rasio,
Mon. Not. Roy. Astron. Soc. \textbf{358} (2005), 572-584
\arXivold{astro-ph/0501131}.

\bibitem{1976A&A....53..259A}
S.~J.~Aarseth and D.~C.~Heggie,
AAP {\bf 53} (1976), 259-265

\bibitem{1971SvA....15..411A}
T.~.A.~Agekyan and Zh.~P.~Anosova,
SovAst {\bf 15} (1971), 411

\bibitem{2010ApJ...717..948I}
N.~Ivanova, S.~Chaichenets, J.~Fregeau, C.~O.~Heinke, J.~C.~Lombardi Jr. and T.~E.~Woods,
ApJ {\bf 717} (2010) no.2, 948-957,
\arXiv{1001.1767}{astro-ph.HE}.

\bibitem{Park:2017zgj}
D.~Park, C.~Kim, H.~M.~Lee, Y.~B.~Bae and K.~Belczynski,
Mon. Not. Roy. Astron. Soc. \textbf{469} (2017) no.4, 4665-4674
\arXiv{1703.01568}{astro-ph.HE}.


\bibitem{KCb}
K.~Kritos and I.~Cholis,
\arXiv{2104.02073}{astro-ph.GA}.

\bibitem{PortegiesZwart:2004ggg}
S.~F.~Portegies Zwart, H.~Baumgardt, P.~Hut, J.~Makino and S.~L.~W.~McMillan,
Nature \textbf{428} (2004), 724
\arXivold{astro-ph/0402622}.

		
\bibitem{H}
D.~C.~Heggie, P.~Hut and S.~L.~W.~McMillan,
Mon. Not. Roy. Astron. Soc. \textbf{318} (2000), L61
\arXivold{astro-ph/9604016}.



\bibitem{Bacon:1996cj}
D.~Bacon, S.~Sigurdsson and M.~B.~Davies,
Mon. Not. Roy. Astron. Soc. \textbf{281} (1996), 830
\arXivold{astro-ph/9603036}.

\bibitem{Antognini:2015ima}
J.~M.~O.~Antognini and T.~A.~Thompson,
Mon. Not. Roy. Astron. Soc. \textbf{456} (2016) no.4, 4219-4246
\arXiv{1507.03593}{astro-ph.SR}.

\bibitem{Fregeau:2004if}
J.~M.~Fregeau, P.~Cheung, S.~F.~Portegies Zwart and F.~A.~Rasio,
Mon. Not. Roy. Astron. Soc. \textbf{352} (2004), 1
\arXivold{astro-ph/0401004}.
%
		
		\bibitem{berti}
		M.~Dominik, E.~Berti, R.~O'Shaughnessy, I.~Mandel, K.~Belczynski, C.~Fryer, D.~E.~Holz, T.~Bulik and F.~Pannarale,
		Astrophys. J. \textbf{806}, no.2, 263 (2015)
		\arXiv{1405.7016}{astro-ph.HE}.
				

	
	\end{thebibliography}
\end{document}